\documentclass[aps, prl, superscriptaddress, nofootinbib, floatfix]{revtex4-2}
\usepackage{amsfonts}
\usepackage{epsfig}
\usepackage{bm}
\usepackage{mathtools, amssymb}
\usepackage{dsfont}
\usepackage{amsmath}
\usepackage{color}
\usepackage{subfigure}
\usepackage{dcolumn}
\usepackage[colorlinks=true, allcolors=magenta]{hyperref}
\usepackage{ulem}

\graphicspath{{figures/}}

\newmuskip\pFqmuskip
\newcommand*\pFq[6][8]{%
  \begingroup 
  \pFqmuskip=#1mu\relax
  \mathchardef\normalcomma=\mathcode`,
  \mathcode`\,=\string"8000
  \begingroup\lccode`\~=`\,
  \lowercase{\endgroup\let~}\pFqcomma
  {}_{#2}F_{#3}{\left(\genfrac..{0pt}{}{#4}{#5}\bigg|#6\right)}%
  \endgroup
}
\newcommand{\pFqcomma}{{\normalcomma}\mskip\pFqmuskip}

\renewcommand{\d}{\mathrm{d}}

\newcommand{\newsection}[1]{{\vspace{10 pt}\noindent \textit{{\textbf{#1}} --}}}

\begin{document}

\title{Divergence and resummation of the moment expansion for an ultrarelativistic gas in Bjorken flow}

\author{Caio V.~P.~de Brito}
\email{caio\_brito@id.uff.br}
\affiliation{Instituto de F\'{\i}sica, Universidade Federal Fluminense \\ Av.~Gal.~Milton Tavares de Souza, S/N, 24210-346, Gragoatá, Niter\'{o}i, Rio de Janeiro, Brazil}
\affiliation{Institute for Theoretical Physics, Goethe University, Max-von-Laue-Str.~1, D-60438 Frankfurt am Main,  Germany}

\author{David Wagner}
\email{david.wagner@unifi.it}
\affiliation{Università degli studi di Firenze,\\
Via G. Sansone 1, I-50019 Sesto Fiorentino (Florence), Italy}

\author{Gabriel S.~Denicol}
\email{gsdenicol@id.uff.br}
\affiliation{Instituto de F\'{\i}sica, Universidade Federal Fluminense \\ Av.~Gal.~Milton Tavares de Souza, S/N, 24210-346, Gragoatá, Niter\'{o}i, Rio de Janeiro, Brazil}

\author{Dirk H. Rischke}
\email{drischke@itp.uni-frankfurt.de}
\affiliation{Institute for Theoretical Physics, Goethe University, Max-von-Laue-Str.~1, D-60438 Frankfurt am Main,  Germany}
\affiliation{Helmholtz Research Academy Hesse for FAIR, Campus Riedberg, 
Max-von-Laue-Str.\ 12, D-60438 Frankfurt am Main, Germany}

\begin{abstract}
In this letter, we demonstrate for the first time that the moment expansion for an ultrarelativistic gas undergoing Bjorken flow diverges. We then show how this series can be resummed using the Borel-Pad\'e method and use this to determine the single-particle distribution function of the gas. Finally, we compare the exact resummed solution of the single-particle distribution function with solutions of the Boltzmann equation in the hydrodynamic limit and verify that the system displays considerable deviations from local equilibrium.
\end{abstract}

\maketitle

\newsection{Introduction} The relativistic Boltzmann equation is widely employed to describe the dynamics of dilute relativistic gases in several areas of physics. 
Applications range from the modeling of the quark-gluon plasma produced in high-energy hadronic collisions \cite{Gale:2013da, Heinz:2013th, Shen:2020mgh, Paquet:2023rfd} to the description of relativistic plasmas in astrophysics \cite{Chandra:2015iza, Most:2021uck}. 
Over the past years, the Boltzmann equation has also been extensively employed to study the emergence of the fluid-dynamical regime in relativistic systems \cite{Denicol:2012cn, Denicol:2012es, Florkowski:2013lya, Denicol:2014mca, Denicol:2014xca, Strickland:2017kux, Florkowski:2017olj, Heller:2020hnq, Bhadury:2022ulr, Wagner:2022ayd, Fotakis:2022usk, deBrito:2023tgb, Rocha:2023hts, Chattopadhyay:2023hpd, Dash:2023ppc, Kushwah:2024zgd, Frasca:2024ege, Strickland:2024moq}.  

There are two widespread methods to derive relativistic fluid dynamics from the Boltzmann equation: the Chapman-Enskog expansion \cite{chapman1970mathematical,cercignani2002} and the method of moments \cite{grad1949, israel1979transient, Denicol:2021}. 
The Chapman-Enskog series is the most traditional formalism to derive fluid dynamics from the Boltzmann equation. 
However, in the relativistic regime, it leads to theories that are acausal and intrinsically unstable when perturbed around global equilibrium \cite{Hiscock:1985zz} and is thus often discarded when studying relativistic gases. 
The method of moments, on the other hand, is a formalism to solve the Boltzmann equation that can also be employed to derive causal and stable \cite{Denicol:2008ha, Pu:2009fj, Brito:2020nou} relativistic fluid-dynamical equations -- the so-called transient theories of fluid dynamics \cite{israel1979transient}.  

The method of moments was proposed by H.~Grad for non-relativistic gases in 1949 \cite{grad1949} and consists in expanding the single-particle distribution function in terms of its moments, using a complete basis in momentum space. 
The Boltzmann equation is then re-expressed as an infinite set of coupled differential equations for these fields \cite{Denicol:2021,deBrito:2024vhm}, which can be in principle solved to reconstruct the single-particle distribution function. 
Relativistic generalizations of Grad's method were first developed by Israel and Stewart in the 1970's \cite{israel1979transient}, using a truncated basis constructed from 4-momentum, and later generalized to a complete basis of irreducible tensors in Ref.~\cite{Denicol:2012cn}.

So far, the moment expansion for the single-particle distribution function has been assumed to converge and its truncations are widely employed in the description of dilute plasmas; for instance the 14-moment approximation has been crucial in modeling the electromagnetic radiation \cite{Paquet:2015lta, Vujanovic:2016anq} and soft hadrons \cite{Molnar:2014fva, JETSCAPE:2020mzn} that are produced in heavy-ion collisions. 
However, the convergence of this series has never been thoroughly investigated.\footnote{The convergence of the moment expansion has been only investigated for homogeneous and isotropic fluids, in the context of cosmology \cite{Bazow:2015dha}.} 
In this letter, we investigate the convergence of the moment expansion for the first time considering a classical gas of massless particles in a simplified description of a heavy-ion collision referred to as Bjorken flow \cite{Bjorken1983}. 
In contrast to common belief, we demonstrate that this expansion actually diverges due to the factorial growth of the moments of the distribution function. 
We then resum this series using the Borel-Pad\'e method \cite{borel1899memoire, pade1892representation} and finally determine the solution of the Boltzmann equation from the method of moments. 
These solutions demonstrate that truncations of the moment expansion often employed in theoretical descriptions of heavy-ion collisions are not reliable approximations.

Throughout this work, we adopt natural units, $c = \hbar = k_B = 1$. 

\newsection{Relativistic kinetic theory in Bjorken flow} We consider a relativistic gas of massless and classical particles undergoing Bjorken flow \cite{Bjorken1983,Baym:1983amj}. 
This flow configuration is an idealized description of a heavy-ion collision that assumes that the matter produced shortly after the collision is homogeneous and azimuthally symmetric in the transverse plane, as well as invariant under Lorentz boosts along the longitudinal axis. 
Furthermore, in Bjorken flow, the otherwise convoluted partial differential equations that govern the evolution of relativistic fluids reduce to simpler ordinary differential equations that admit analytical solutions \cite{deBrito:2024vhm}. 
Hence, Bjorken flow often serves as a starting point in studies of relativistic kinetic theory \cite{Heller:2015dha, Heinz:2015gka, Chen:2023vrk, Nugara:2023eku, deBrito:2024vhm, Frasca:2024ege} and fluid dynamics \cite{Jaiswal:2021uvv, Dash:2022xkz, deBrito:2023tgb, Chattopadhyay:2023hpd}.

The Bjorken flow profile is more conveniently described using hyperbolic rather than Cartesian coordinates, replacing the time and $z$ coordinates by the \textit{proper} time, $\tau = \sqrt{t^2-z^2}$, and spacetime rapidity, $\eta_s = \mathrm{Artanh} (z/t)$, respectively. The corresponding metric tensor is then given by $g_{\mu\nu} = \mathrm{diag}(1,-1,-1,-\tau^2)$. 
In this case, the Boltzmann equation reads 
\begin{equation}
\label{eq:boltz-eq}
 \partial_\tau f_{\mathbf{k}}= \frac{1}{E_\mathbf{k}} C[f_{\mathbf{k}}] \approx - \frac{1}{\tau_R} \left( f_{\mathbf{k}} - f_{0 \mathbf{k}} \right)\;,
\end{equation}
where $f_{0\mathbf{k}} = e^{\alpha -\beta E_{\mathbf{k}}}$ is the Boltzmann distribution function in equilibrium, with $\alpha$ being the thermal potential, $\beta$ the inverse temperature, and $E_{\mathbf{k}}$ the energy of the particle. 
For the sake of simplicity, we employed the relaxation-time approximation for the collision term \cite{Anderson1974,Rocha:2021zcw}, where the parameter $\tau_R$ dictates the time scale over which the single-particle distribution function relaxes to its equilibrium value. 
The relaxation time can be expressed in terms of the shear viscosity, $\eta$, as \cite{Denicol:2012cn},
\begin{equation}
\label{eq:shear-relax-time}
\tau_R = \frac{5 \eta}{\varepsilon + P}\;,
\end{equation}
where $\varepsilon$ is the energy density and $P$ is the thermodynamic pressure. 
In the following, we assume that the shear viscosity over entropy density, $\eta/s$, is constant.

Following the method of moments, we now expand the single-particle distribution using a complete basis in momentum space -- in our case, we employ Legendre, $P_n$, and associated Laguerre, $L_n^{(m)}$, polynomials. The result is \cite{Denicol:2021, deBrito:2024vhm},
\begin{equation}
\label{eq:dist_func}
f_{\mathbf{k}}
= 
f_{0\mathbf{k}} \sum_{\ell = 0}^\infty \sum_{n = 0}^\infty c_{n, \ell} (\beta E_{\mathbf{k}} )^{2\ell}  P_{2\ell}(\cos\Theta) L_n^{(4\ell + 1)}(\beta E_{\mathbf{k}})\;,
\end{equation}
where we defined the angle $\Theta$ in momentum space via $\cos\Theta \equiv k_{\eta_s}/(\tau E_{\mathbf{k}})$, with $k_{\eta_s}$ being the longitudinal component of the 4-momentum. 
The expansion coefficients $c_{n, \ell}$ are determined using the orthogonality relations satisfied by the Legendre and associated Laguerre  polynomials, leading to
\begin{equation}
\label{eq:exp-coeffs-rho}
c_{n, \ell}
=
(4\ell + 1) n! \sum_{m=0}^n \frac{(-1)^m \, (m + 2\ell + 1)!}{(n-m)! (m + 4\ell + 1)! m!} \,\chi_{m + 2\ell, \ell}\;.
\end{equation}
Here, we defined $\chi_{n, \ell} = \varrho_{n, \ell}/\varrho^{\mathrm{eq}}_{n, 0}$, where $\varrho_{n, \ell}$ are the irreducible moments of a general distribution function $f_{\mathbf{k}}$ in Bjorken flow,
\begin{equation}
\label{eq:def-rho-bjork}
\varrho_{m, n} = \int \mathrm{d}K E_{\mathbf{k}}^{m} P_{2n}(\cos\Theta) f_{\mathbf{k}}\;,
\quad \text{and} \quad
\varrho_{n, \ell}^{\mathrm{eq}} = e^\alpha \frac{(n+1)!}{2 \pi^2 \beta^{n+2}} \, \delta_{\ell 0}\;,
\end{equation}
with $\mathrm{d}K=\mathrm{d}^3 \mathbf{k}/\left[(2\pi)^3 E_{\mathbf{k}} \tau \right]$. 
Recursively using the Boltzmann equation \eqref{eq:boltz-eq}, one can derive a set of coupled ordinary differential equations for the irreducible moments, cf.~Refs.~\cite{Denicol:2021, deBrito:2024vhm}. 
In order to work with dimensionless variables, it is convenient to obtain a hierarchy of differential equations for the rescaled moments $\chi_{n, \ell}$ as a function of the \textit{normalized} proper time, $\hat{\tau} \equiv \tau/\tau_R$.
We remark that with this choice of variables, the dynamics of the (rescaled) fields no longer explicitly depends on the viscosity and more dissipative systems can be studied by considering smaller values of the initial rescaled time. 
The rescaled moments then satisfy the following equations of motion (see Refs.~\cite{Denicol:2021, deBrito:2024vhm} for details)
\begin{equation}
\frac{\mathrm{d} \chi_{n, \ell}}{\mathrm{d} \hat{\tau}}
=
\frac{3}{2} \left[ 1 - \frac{\chi_{2,1} (1 - \alpha)}{4-\alpha} \right]^{-1} \left[ -  \chi_{n, \ell} + \delta_{\ell 0}  - \mathcal{P}_{n, \ell} \frac{\chi_{n, \ell - 1}}{\hat{\tau}} - \mathcal{Q}_{n, \ell} \frac{\chi_{n, \ell}}{\hat{\tau}} - \mathcal{R}_{n, \ell} \frac{\chi_{n, \ell + 1}}{\hat{\tau}} + \frac{2 (n-1)}{3 \hat{\tau}} \chi_{2,1} \chi_{n, \ell} \right]\;, \label{eq:resc-eoms-moms}
\end{equation}
where we have introduced the following coefficients
\begin{equation}
\mathcal{P}_{n, \ell} = \frac{2 \ell (n + 2\ell ) (2\ell - 1)}{(4\ell + 1) (4\ell - 1)}\;, \quad
\mathcal{Q}_{n, \ell} = \frac{2 \ell (2\ell + 1) (2n + 1)}{3 (4\ell - 1) (4\ell + 3)}\;, \quad
\mathcal{R}_{n, \ell} = (n - 2 \ell -1) \frac{(2\ell + 1) (2\ell + 2)}{(4\ell + 1) (4\ell + 3)}\;.
\end{equation}
Moments of a given rank are always coupled with those of higher and lower ranks, given by $\chi_{n, \ell+1}$ and $\chi_{n, \ell-1}$, respectively. 
Meanwhile, moments with different powers of energy do not couple with each other, a particular feature of the relaxation-time approximation for massless particles \cite{Denicol:2021}.

As usual, we employ Landau matching conditions \cite{landau:59fluid} to define the thermal potential and temperature out of equilibrium,\footnote{A generalization for the relaxation-time approximation in which different choices of matching conditions do not violate the conservation laws has been addressed in Ref.~\cite{Rocha:2021zcw}.} in which case the particle and energy density are always fixed to their corresponding equilibrium values,
\begin{eqnarray}
n \equiv n_0 \Longrightarrow \varrho_{1,0} = \varrho_{1,0}^{\mathrm{eq}}\;, 
\quad
\varepsilon \equiv \varepsilon_0 \Longrightarrow \varrho_{2,0} = \varrho_{2,0}^{\mathrm{eq}}\;,
\end{eqnarray}
which implies that $\chi_{1,0} = \chi_{2,0} = 1$. Then, the thermal potential and temperature are obtained from the conservation laws, which can be rewritten as,
\begin{equation}
\label{eq:thermo-eoms}
\frac{\mathrm{d} \alpha}{\mathrm{d} \hat{\tau}} - \frac{3}{2} \left[ 1 - \frac{\chi_{2,1} (1 - \alpha)}{4-\alpha} \right]^{-1} \left(  \frac{2 \chi_{2,1}}{\hat{\tau}} \right) = 0\;, \quad
\frac{\mathrm{d}T}{\mathrm{d} \hat{\tau}} + \frac{3}{2} \left[ 1 - \frac{\chi_{2,1} (1 - \alpha)}{4-\alpha} \right]^{-1} \frac{T}{3 \hat\tau} (1 + 2 \chi_{2,1}) = 0\;. 
\end{equation}

Once the moment equations are solved, it is then possible to reconstruct the single-particle distribution function using Eq.\ \eqref{eq:dist_func}. 
In principle, an \textit{exact} solution for the distribution function would only be obtained with the inclusion of an infinite number of moments in the moment expansion \eqref{eq:dist_func}. 
A priori, we expect that the inclusion of more moments in the moment expansion \eqref{eq:dist_func} will lead to a more accurate solution for the single-particle distribution function. 

\newsection{Multipoles of the distribution function} The main goal of this letter is to compute, for the first time, the single-particle distribution function of a gas via the moment expansion. We first re-express Eq.~\eqref{eq:dist_func} as,
\begin{equation}
\label{eq:multipole}
f_{\mathbf{k}}
=
\sum_{\ell=0}^{\infty} \frac{4\ell+1}{2} P_{2\ell}( \cos\Theta ) \mathcal{F}_{\mathbf{k}}^{( \ell )},
\end{equation}
where we introduced the $\ell$-th multipole coefficient
\begin{equation}
\label{eq:multipoles-fk}
\mathcal{F}^{(\ell)}_{\mathbf{k}} 
\equiv \int_{-1}^{1} \mathrm{d}\cos\Theta P_{2\ell}( \cos\Theta ) \ f_{\mathbf{k}} 
= \frac{2}{4\ell+1} (\beta E_{\mathbf{k}})^{2\ell} f_{0\mathbf{k}} \sum_{n = 0}^\infty c_{n, \ell} L_n^{(4\ell + 1)}(\beta E_{\mathbf{k}})\;.
\end{equation}

\begin{figure}[ht]
\centering
\includegraphics[width=\linewidth]{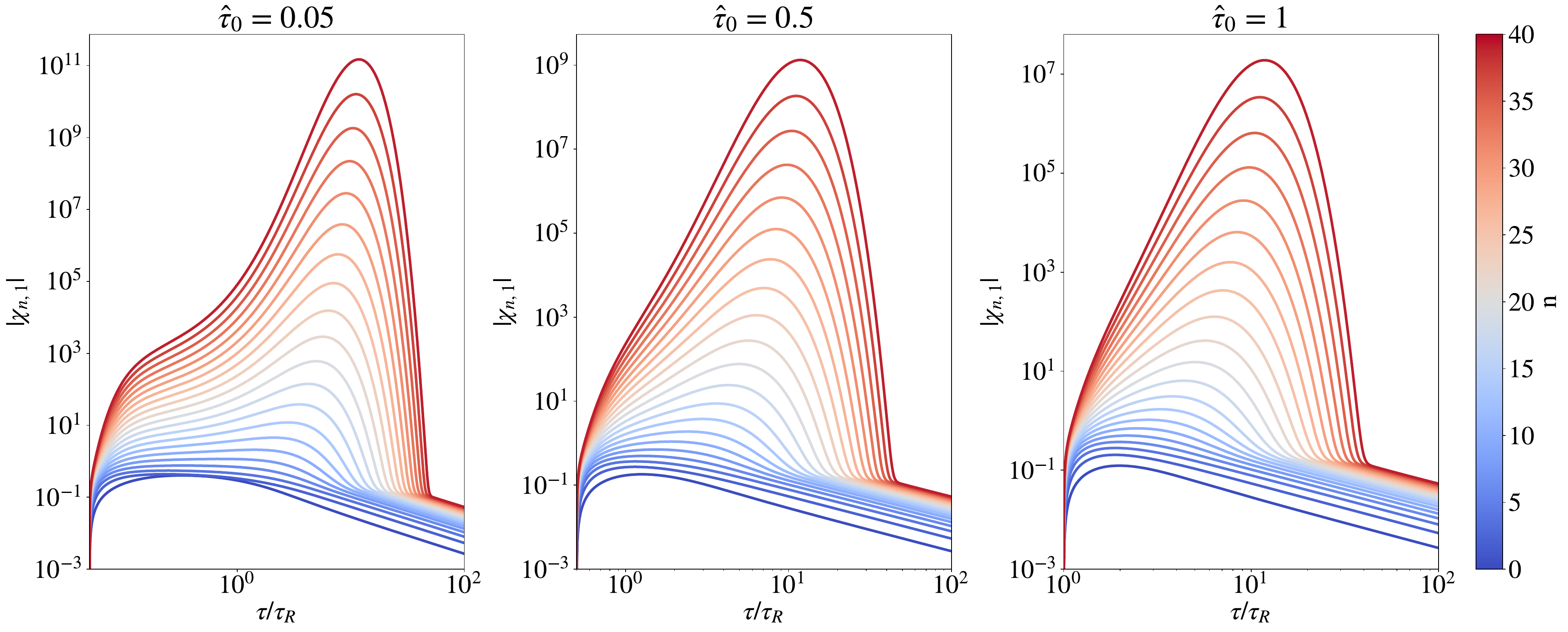}
\caption{Irreducible moments normalized to their equilibrium values as a function of the rescaled proper time considering $\hat{\tau}_0 = 0.05$ (left panel), $0.5$ (middle panel), and $1$ (right panel). 
In all three cases, we assume equilibrium initial conditions, $\chi_{n \ell} = \delta_{\ell 0}$, $T(\hat{\tau}_0) = 10$ GeV, and $n(\hat{\tau}_0) = T(\hat{\tau}_0)^3/\pi^2$.}
\label{fig:resc-moments}
\end{figure}

First, we analyze the solutions for the normalized irreducible moments $\chi_{n, \ell}$. 
We solve Eq.~\eqref{eq:resc-eoms-moms} for values of $n$ ranging from $n=0, \ldots, 40$ and $\ell=0, \ldots, 20$, coupled with Eqs.~\eqref{eq:thermo-eoms}. 
We consider three values for the initial rescaled proper time, namely $\hat{\tau}_0 = 0.05$, $0.5$, and $1$, and assume equilibrium initial conditions, $\chi_{n, \ell} \ (\hat{\tau}_0) = \delta_{\ell 0}$. 
In Fig.~\ref{fig:resc-moments}, we portray the rescaled moments $\chi_{n1}$ as functions of the rescaled proper time for a wide range of $n$. 
We observe that the irreducible moments increase significantly in magnitude as $n$ increases, with this behavior becoming more pronounced for smaller values of $\hat{\tau}_0$, that is, for more dissipative systems. We remark that moments of different ranks (here denoted by $\ell$) also display the same qualitative behavior. 
In the supplemental material, we demonstrate analytically (for a constant relaxation time) that the moments $\chi_{n\ell}$ actually grow factorially. 
Thus, the expansion of the multipole coefficients in terms of Laguerre polynomials \eqref{eq:multipoles-fk} tends to be problematic.
On the other hand, we note that, for a fixed $n$, the moments \textit{decrease} in magnitude as $\ell$ is increased. 
Thus, the multipole expansion for $f_{\mathbf{k}}$, given in Eq.~\eqref{eq:multipole}, is expected to converge rapidly. 

\begin{figure}[ht]
\centering
\includegraphics[width=\linewidth]{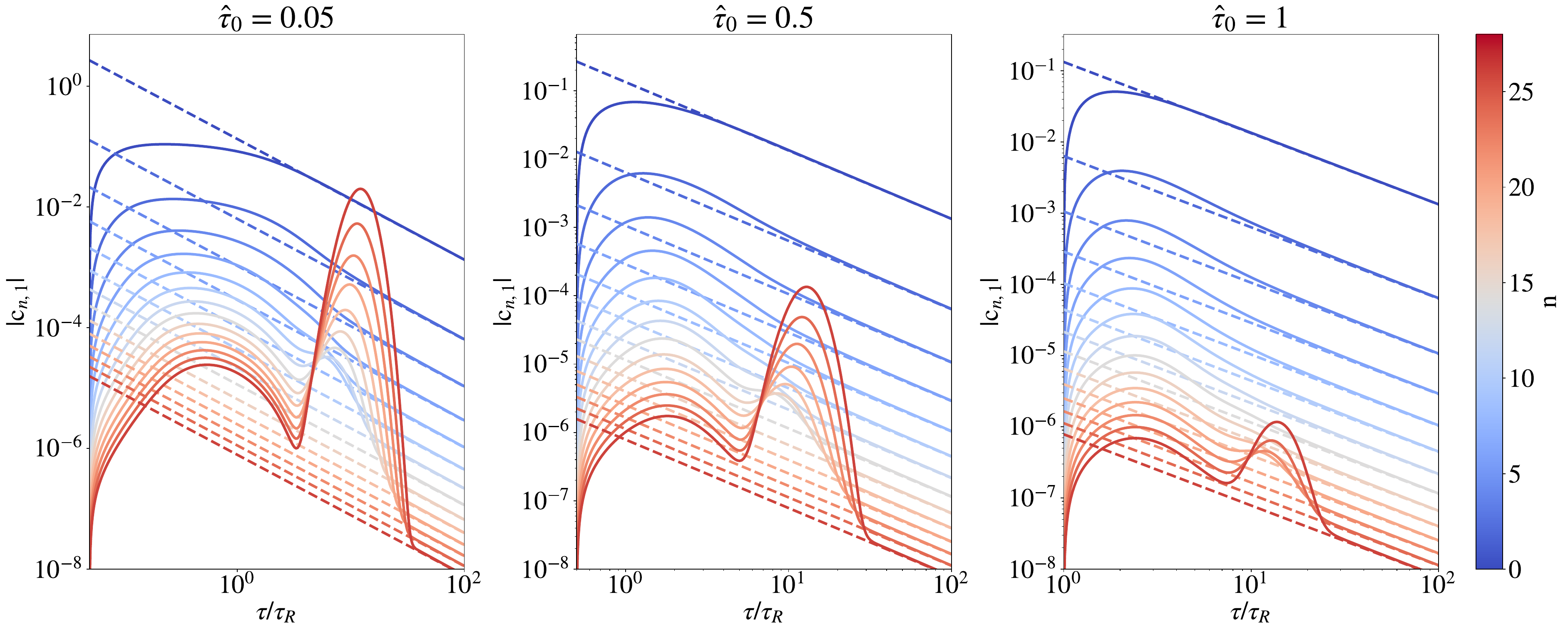}
\caption{Exact solutions (solid lines) and Navier-Stokes limit (dashed lines) for the expansion coefficients as a function of the rescaled proper time considering $\hat{\tau}_0 = 0.05$ (left panel), $0.5$ (middle panel), and $1$ (right panel). 
In all three cases, we assume equilibrium initial conditions, $\chi_{n \ell} = \delta_{\ell 0}$, $T(\hat{\tau}_0) = 10$ GeV, and $n(\hat{\tau}_0) = T(\hat{\tau}_0)^3/\pi^2$.}
\label{fig:expn-coeffs}
\end{figure}

The divergence of the series can be observed using Fig.~\ref{fig:expn-coeffs}, where we display the expansion coefficients $c_{n, 1}$ (used to reconstruct the multipole $\mathcal{F}^{(1)}_{\mathbf{k}}$) as a function of the rescaled time. 
At both early and late times, we see that the expansion coefficients decrease with $n$. 
However, for intermediate values of the rescaled proper time, the expansion coefficients display a qualitatively different behavior and start to increase with $n$, with this behavior becoming more pronounced for smaller values of the \textit{initial} rescaled time. 
We remark that, in order for the expansion \eqref{eq:multipoles-fk} in associated Laguerre polynomials to converge, the expansion coefficients $c_{n, 1}$ must not increase with $n$ and, thus, the aforementioned behavior guarantees that Eq.~\eqref{eq:multipoles-fk} diverges at least for these intermediate values of the rescaled time. 
A qualitatively similar behavior is also observed for other values of $\ell$. 

\begin{figure}[h]
\centering
\includegraphics[width=\linewidth]{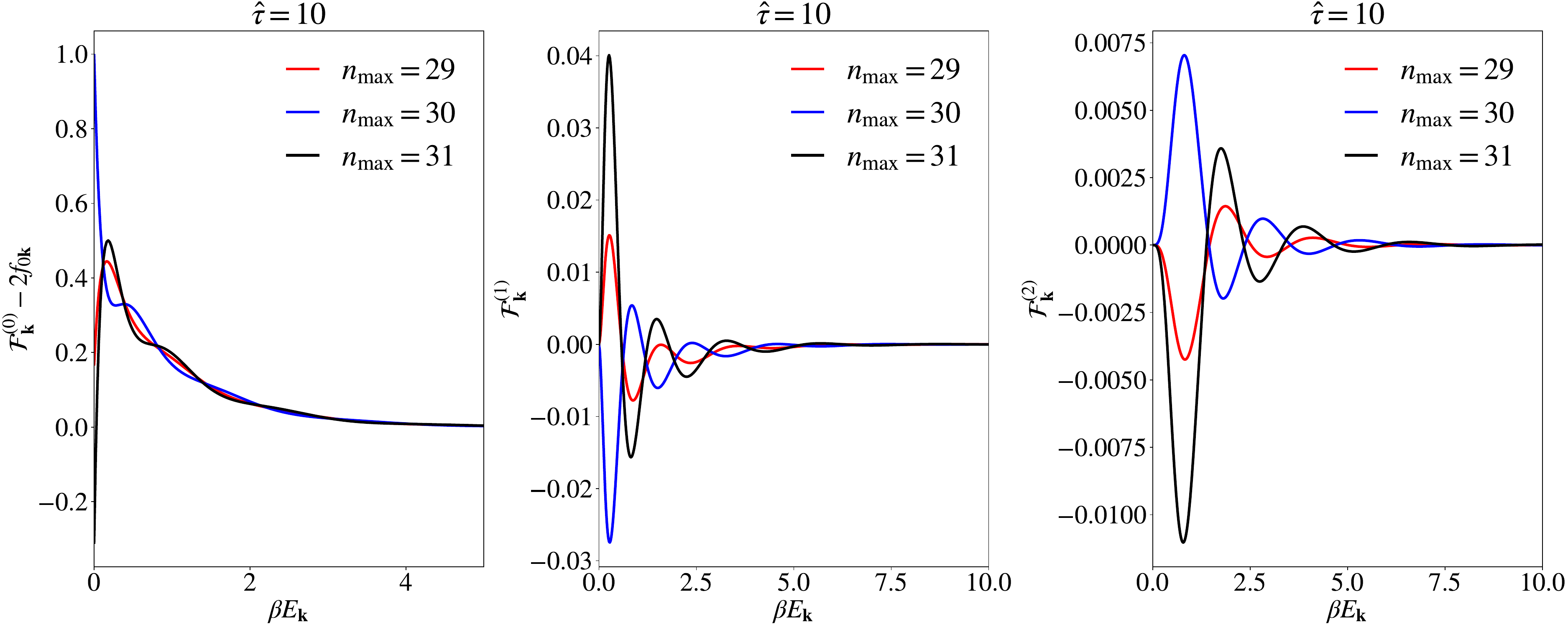}
\caption{Multipoles of the single-particle distribution function as functions of $\beta E_{\mathbf{k}}$, for different truncations, considering $\hat{\tau}_0 = 0.5$ and $\hat\tau = 10$. }
\label{fig:diverg-multipoles}
\end{figure}

We explicitly verify the non-convergence of the moment expansion in Fig.~\ref{fig:diverg-multipoles}, where we compute a snapshot ($\hat\tau = 10$) of the first three multipoles of $f_{\mathbf{k}}$, for $\hat{\tau}_0 = 0.5$, as a function of $\beta E_{\mathbf{k}}$ and considering three different truncations of the series -- the parameter $n_{\mathrm{max}}$ specifies the number of terms included in the series \eqref{eq:multipoles-fk}. 
As expected, we indeed observe that the expansion for the three multipoles portrayed does not converge, exhibiting an unphysical oscillatory pattern as more terms are included in the moment expansion. 
Hence, this implies that the expansion of the single-particle distribution itself is also divergent. 
In the supplemental material we demonstrate analytically that the moment expansion for the distribution function diverges for rescaled times of $\hat{\tau}/\hat{\tau}_0 > $ 8 in the case of a constant relaxation time.

\newsection{Resummed moment expansion} We have shown that the expansion of the distribution function \eqref{eq:multipoles-fk} diverges for a wide range of values of the rescaled proper time. 
We now evaluate this series using the Borel-Padé resummation scheme. 
We first calculate the Borel transform \cite{borel1899memoire} of Eq.~\eqref{eq:multipoles-fk}, leading to
\begin{eqnarray}
\label{eq:borel-transf-series}
\mathcal{F}^{(\ell)}_{\mathbf{k}} 
= \frac{2}{4\ell+1} (\beta E_{\mathbf{k}})^{2\ell} f_{0\mathbf{k}} \int_0^\infty \mathrm{d}y \ e^{-y} \sum_{n = 0}^\infty \frac{y^n}{n!} c_{n, \ell} L_n^{(4\ell + 1)}(\beta E_{\mathbf{k}})\;.
\end{eqnarray}
Then, we re-express the series $f(y) = \sum_{n = 0}^\infty \frac{y^n}{n!} c_{n, \ell} L_n^{(4\ell + 1 )}(\beta E_{\mathbf{k}})$ using Padé approximants \cite{pade1892representation} and calculate the integral in Eq.~\eqref{eq:borel-transf-series} numerically. 
The Padé approximant of $f(y)$ is defined as \cite{Baker:1996}
\begin{equation}
f^{[q,s]} (y) = \frac{\sum_{i=0}^q a_i y^i}{\sum_{j=0}^s b_j y^j}\;.
\end{equation}
Without loss of generality, we take $b_0 = 1$, which effectively reduces the number of unknown coefficients to $q+s+1$. 
The coefficients $a_i$ and $b_i$ are determined so that $f^{[q,s]} (y)$ matches the original series up to order $\mathcal{O}(y^{q+s})$. 
In particular, we have the freedom to adjust the coefficients $q$ and $s$ to obtain the best fit to the resummed series.\footnote{In the Supplemental Material, we show that the Padé approximants capture the optimal truncation of the Borel-transformed series for $\mathcal{F}_{\mathbf{k}}^{(\ell)}$.}

\begin{figure}[h]
\centering
\includegraphics[width=\linewidth]{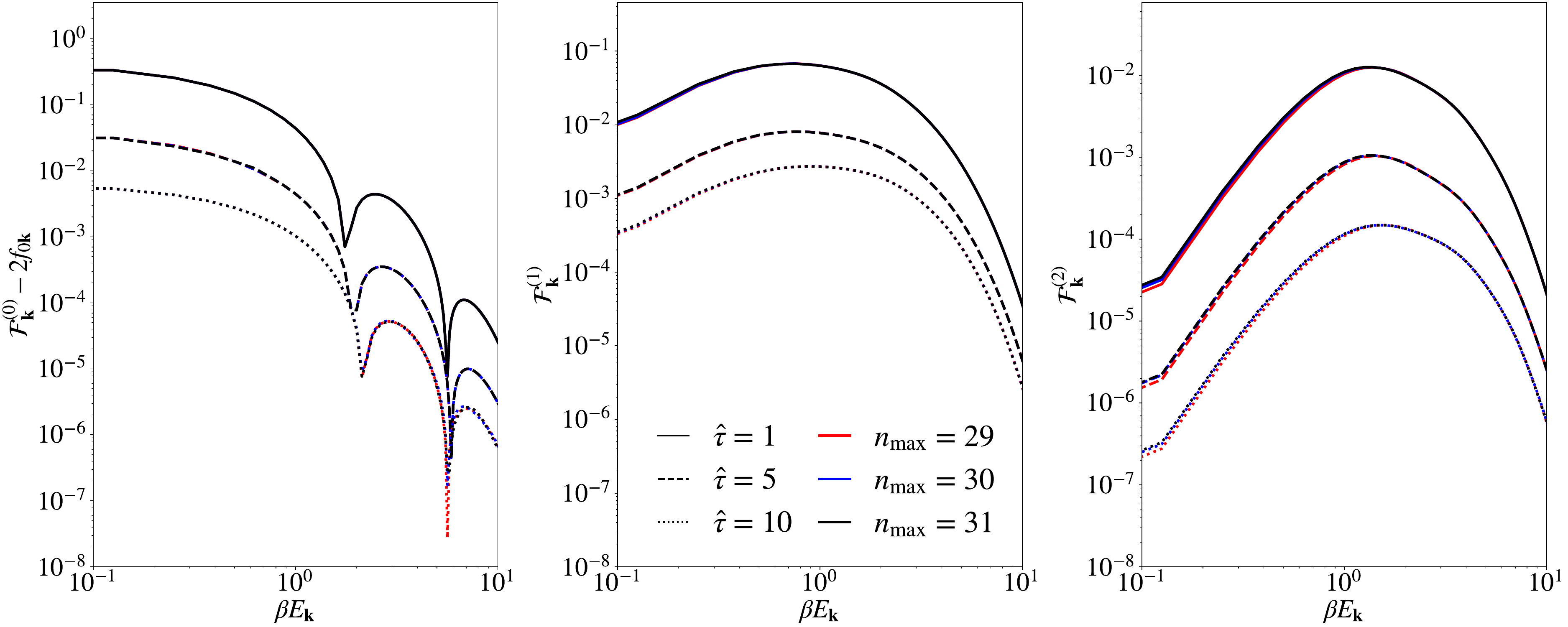}
\caption{Multipoles of the single-particle distribution function as functions of $\beta E_{\mathbf{k}}$ for different truncations, computed with a Borel-Padé resummation scheme, for $\hat\tau = 1, 5, 10$ considering $\hat{\tau}_0 = 0.5$.}
\label{fig:resum-multipoles}
\end{figure}

In Fig.~\ref{fig:resum-multipoles}, we show the first three multipoles of the distribution function as a function of $\beta E_{\mathbf{k}}$ for $\hat{\tau}_0 = 0.5$, computed using a Borel-Padé resummation scheme, considering the same truncations as in Fig.~\ref{fig:diverg-multipoles}. 
We display results for three different times, namely $\hat\tau = 1$ (solid lines), $\hat\tau = 5$ (dashed lines), and $\hat\tau = 10$ (dotted lines), for which it can be readily seen that all multipoles depicted display convergent behavior.
We remark that in computing the zeroth ($\ell=0$) and second ($\ell=2$) multipoles, we have fixed the order of the polynomial in the denominator of the Padé approximant to be $s=2$, whereas for the first ($\ell=1$), we take $s=4$, with the order of the corresponding numerator being computed accordingly in each case.
This choice is made to avoid the occurrence of spurious poles -- pseudo-singularities that do not exist in the original function -- in the real plane, which ultimately lead to numerical errors.

\begin{figure}[h]
\centering
\includegraphics[width=\linewidth]{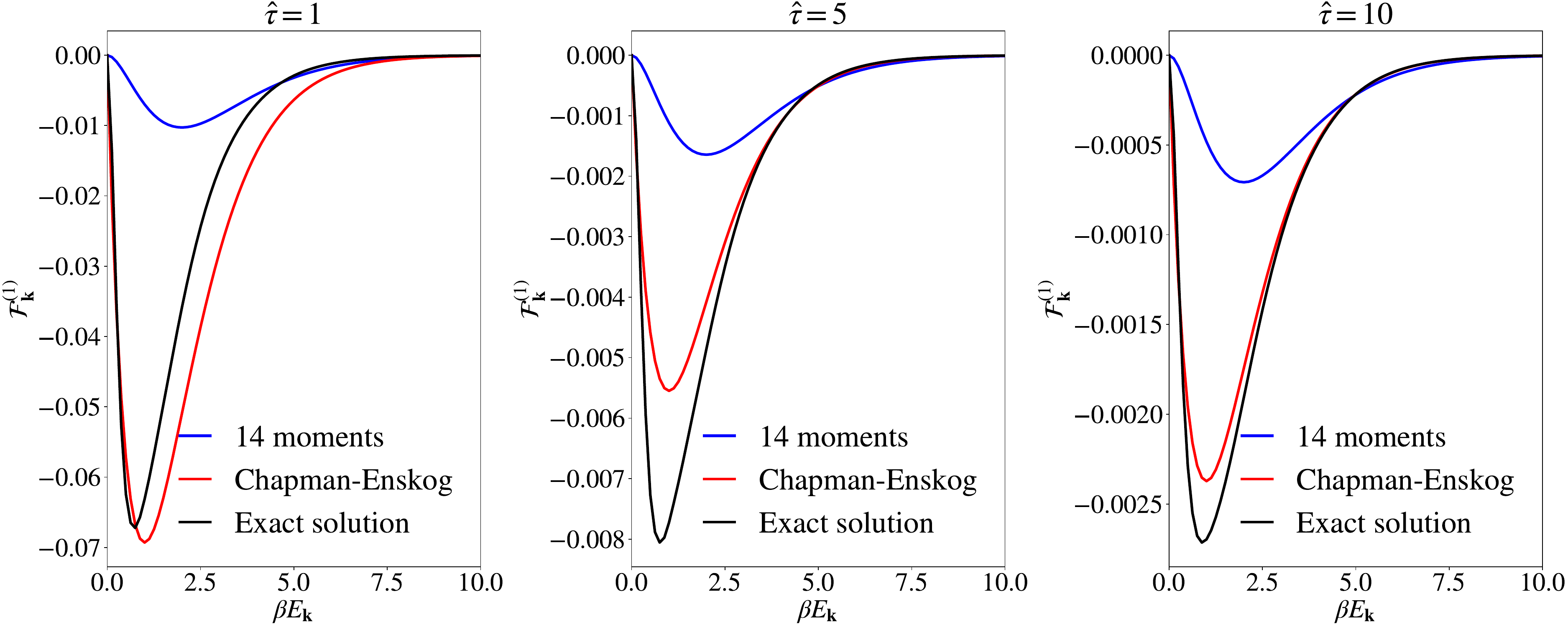}
\caption{Comparison of the exact and hydrodynamic solutions with the Chapman-Enskog limit of $\mathcal{F}_{\mathbf{k}}^{(1)}$ as functions of $\beta E_{\mathbf{k}}$ for $\hat\tau = 1, 5, 10$ considering $\hat{\tau}_0 = 0.5$.}
\label{fig:comparison-f}
\end{figure}

We also compare the exact solutions for the multipoles with traditional approximations that are expected to be valid in the fluid-dynamical limit: the first-order Chapman-Enskog solution \cite{cercignani2002} and the 14-moment approximation \cite{israel1979transient}. 
Since both of these hydrodynamic solutions only contain the $\ell = 1$ multipole of the single-particle distribution function, we restrict our comparison to this case. 
In Fig.~\ref{fig:comparison-f}, we plot $\mathcal{F}^{(1)}_{\mathbf{k}}$ as a function of $\beta E_{\mathbf{k}}$ considering $\hat{\tau}_0 = 0.5$. 
We observe that the 14-moment approximation, traditionally employed in fluid-dynamical models of heavy-ion collisions, is not in good agreement with the exact solution for $\mathcal{F}^{(1)}_{\mathbf{k}}$.
On the other hand, the first-order Chapman-Enskog solution is in better agreement with the exact solution, in particular at late times. 
In Fig.~\ref{fig:resum-multipoles} we can also see that, at early times, the $\ell=0$ and $\ell=2$ multipole components, which should vanish in the fluid-dynamical limit, are comparable in magnitude to the $\ell=1$ component. 
This further indicates a significant deviation of the solution of the Boltzmann equation from the fluid-dynamical limit. 

\newsection{Conclusions} In this work we computed the single-particle distribution function for an expanding classical massless gas using the method of moments. 
For this purpose, we expanded the distribution function in a basis of Legendre and associated Laguerre polynomials and converted the Boltzmann equation into an infinite set of coupled differential equations for its irreducible moments. 
We then solved these equations of motion in order to reconstruct the single-particle distribution function.

We demonstrated for the first time that the moment expansion for an ultra-relativistic gas undergoing Bjorken flow diverges. 
We then showed how this series can be resummed using the Borel-Pad\'e method and determined the single-particle distribution function. 
We finally compared the exact resummed solution of the single-particle distribution function with solutions of the Boltzmann equation in the hydrodynamic limit: the 14-moment approximation and the first-order Chapman-Enskog solution. 
We observed that these hydrodynamic solutions are not in good agreement with the exact solution, demonstrating that this system is always far from a hydrodynamic regime.

Finally, we remark that the divergence of the moment expansion will have a significant impact on the derivation of transient fluid dynamics from the Boltzmann equation, where the convergence of this series and its truncations were thoroughly employed. 
Our results will also have an impact in the fluid-dynamical models of heavy-ion collisions, where truncated moment expansions are often employed to approximate the momentum distribution of partons and hadrons throughout the collision process. 
These issues will be addressed in future works.  

\newsection{Acknowledgments}  C.V.P.B.~acknowledges the hospitality of the Institute for Theoretical Physics of Goethe University, where part of this work was done. 
C.V.P.B.~is partly funded by Coordenação de Aperfeiçoamento de Pessoal de Nível Superior (CAPES), Finance Code 001, Award No. 88881.722616/2022-01 and by Conselho Nacional de Desenvolvimento Científico e Tecnológico (CNPq), Grant No. 140453/2021-0. G.S.D.~also acknowledges CNPq as well as Fundação Carlos Chagas Filho de Amparo à Pesquisa do Estado do Rio de Janeiro (FAPERJ), Grant No.~E-26/202.747/2018. D.W.~acknowledges support by the project PRIN2022 Advanced Probes of the Quark Gluon Plasma funded by ”Ministero dell’Università e della Ricerca”.
D.W.~and D.H.R.~acknowledge support by the Deutsche Forschungsgemeinschaft (DFG, German Research Foundation) through the CRC-TR 211 ``Strong-interaction matter under extreme conditions'' -- project number 315477589 -- TRR 211, and by the State of Hesse within the Research Cluster ELEMENTS (Project ID 500/10.006).

\bibliographystyle{apsrev4-1} 
\bibliography{refs}


\newpage

\begin{center}
  \textbf{\large Divergence and resummation of the moment expansion for an ultrarelativistic gas in Bjorken flow\\Supplemental Material} \\[.2cm]
  Caio V. P. de Brito,$^{1, 2}$ David Wagner,$^3$ Gabriel S. Denicol,$^1$ and Dirk H. Rischke$^{2, 4}$\\[.1cm]
{\itshape ${}^1$Instituto de Física, Universidade Federal Fluminense \\
Av. Gal. Milton Tavares de Souza, S/N, 24210-346, Gragoatá, Niterói, Rio de Janeiro, Brazil\\}
{\itshape ${}^2$Institute for Theoretical Physics, Goethe University, \\
Max-von-Laue-Str. 1, D-60438 Frankfurt am Main, Germany\\}
{\itshape ${}^3$Università degli studi di Firenze, \\
Via G. Sansone 1, I-50019 Sesto Fiorentino (Florence), Italy\\}
{\itshape ${}^4$Helmholtz Research Academy Hesse for FAIR, Campus Riedberg, \\
Max-von-Laue-Str. 12, D-60438 Frankfurt am Main, Germany\\}
\end{center}

In this supplemental material, we provide further notes on the Borel-Padé resummation employed in the main text and present an analytical approach to elucidate the origin of the observed divergence of the moment expansion.

\section{Multipole expansion of the single-particle distribution function}
\label{sec:mult_exp}

The main goal was to compute, for the first time, the \textit{complete} single-particle distribution function for a classical gas of massless particles in Bjorken flow, in which case it can be expanded as
\begin{equation}
f_{\mathbf{k}}
=
f_{0\mathbf{k}} \sum_{\ell = 0}^\infty \sum_{n = 0}^\infty (\beta E_{\mathbf{k}})^{2\ell} c_{n, \ell} L_n^{(4\ell + 1)}(\beta E_{\mathbf{k}}) P_{2\ell}(\cos\Theta)\;, 
\end{equation}
where the expansion coefficients $c_{n, \ell}$ are defined as
\begin{equation}
c_{n, \ell} = (4\ell + 1) n! \sum_{m=0}^n \frac{(-1)^m \, (m + 2\ell + 1)!}{(n-m)! (m + 4\ell + 1)! m!} \chi_{m + 2\ell, \ell}\;.
\end{equation}

For the sake of convenience, hereon we shall work with integrals of $f_{\mathbf{k}}$ rather than with the complete distribution function itself.
For this purpose, we define the $\ell$-th multipole of $f_{\mathbf{k}}$ as
\begin{equation}
\mathcal{F}^{(\ell)}_{\mathbf{k}} 
\equiv \int_{-1}^{1} \mathrm{d}\cos\Theta \, P_{2\ell}( \cos\Theta ) \ f_{\mathbf{k}} 
= \frac{2}{4\ell+1} (\beta E_{\mathbf{k}})^{2\ell} f_{0\mathbf{k}} \sum_{n = 0}^\infty c_{n, \ell} L_n^{(4\ell + 1 )}(\beta E_{\mathbf{k}})\;.
\end{equation}
Hence, the distribution function can be expressed as
\begin{equation}
f_{\mathbf{k}}
=
\sum_{\ell=0}^{\infty} \frac{4\ell+1}{2} P_{2\ell}( \cos\Theta ) \mathcal{F}_{\mathbf{k}}^{( \ell )}\;.
\end{equation}

In the main text, we showed that this is, in fact, a divergent series. In particular, this implies that the non-equilibrium distribution function is not square-integrable. Therefore, in order to assess physically meaningful results from the moment expansion of the single-particle distribution function, it is necessary to resort to resummation schemes.

\subsection{Borel transform}

\begin{figure}[h]
\centering
\includegraphics[width=\linewidth]{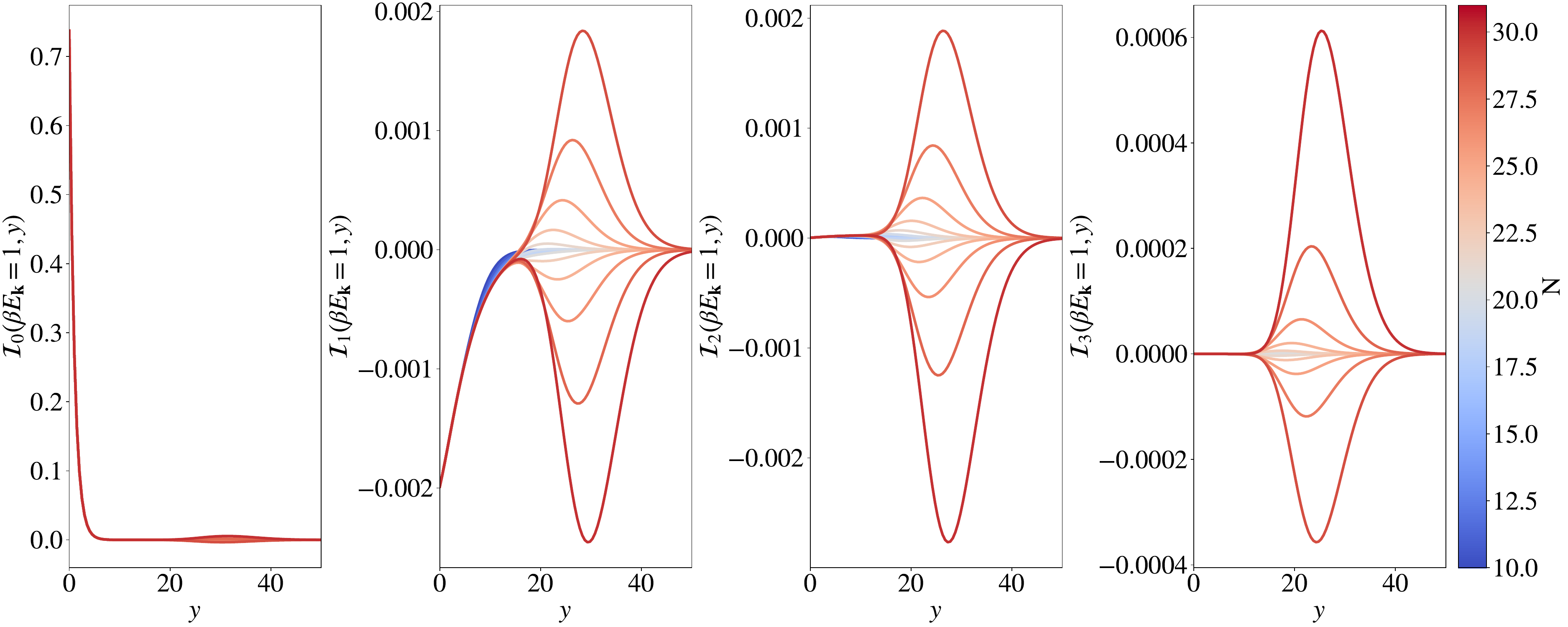}
\caption{Integrand of the Borel-transformed series of different multipoles for a wide range of truncations $N$, considering $\beta E_{\mathbf{k}} = 1$, $\hat{\tau}_{0} = 0.5$ and $\hat{\tau}=10$.}
\label{fig:integrand-borel}
\end{figure}

In particular, the factorial divergence of the normalized irreducible moments -- analytically shown considering a simplified version of the hierarchy of moment equations -- suggests that a Borel resummation might lead to a convergent series. 
We then perform a Borel transform in the sum over $n$, leading to
\begin{equation}
\label{eq:app-borel-transf-series}
\mathcal{F}_{\mathbf{k}}^{( \ell )}
=
\frac{2}{4\ell+1} ( \beta E_{\mathbf{k}})^{2\ell} f_{0\mathbf{k}}  \int_0^\infty \mathrm{d} y \, e^{-y} \underbrace{\sum_{n = 0}^{N} \frac{y^n}{n!} c_{n, \ell}  L_n^{(4\ell + 1 )}(\beta E_{\mathbf{k}})}_\text{Borel-transformed series}\;.
\end{equation}
In particular, let us analyze the integrand of the equation above,
\begin{equation}
\label{eq:integrand-borel}
\mathcal{I}_{\ell}(\beta E_{\mathbf{k}}, y ) \equiv \frac{2}{4\ell+1} (\beta E_{\mathbf{k}})^{2\ell} f_{0\mathbf{k}} \, e^{-y} \sum_{n = 0}^{N} \frac{y^n}{n!} c_{n, \ell}  L_n^{(4\ell + 1)}(\beta E_{\mathbf{k}})\;.
\
\end{equation}
In Fig.~\ref{fig:integrand-borel}, we portray $\mathcal{I}_{\ell}( \beta E_{\mathbf{k}}=1, y )$ as a function of $y$ for a wide range of different truncations of the sum in Eq.~\eqref{eq:integrand-borel}. 
From this plot, it can be straightforwardly seen that there exists an optimal truncation $N$ for the Borel transform of the multipoles, beyond which the results for $\mathcal{I}_{\ell}$ start to diverge in the range $20 \lesssim y \lesssim 40$. This is a clear signature of a divergent series, and thus serves as another striking evidence of such pathological behavior of the moment expansion in Bjorken flow. Therefore, we conclude that, in order to obtain a convergent series, it is necessary to employ further resummation techniques. 

\subsection{Padé approximants}

We have shown that multipoles of the single-particle distribution function, and, thus, $f_{\mathbf{k}}$, are not Borel-resummable. 
Therefore, we shall try computing the Borel transform of the multipoles, Eq.~\eqref{eq:app-borel-transf-series}, using Padé approximants, a rational representation of a function whose Maclaurin expansion recovers the power-series representation of the original function up to a given order \cite{Baker:1996}. 
The Padé approximant of a function that can be expressed as $f(x) = \sum_{i=0}^\infty c_i x^i$ is defined as
\begin{equation}
f(x) = \sum_{i=0}^\infty c_i x^i \quad \Longrightarrow \quad f^{[q,s]} (x) = \frac{\sum_{i=0}^q a_i x^i}{\sum_{j=0}^s b_j x^j} \quad \therefore \quad f(x) = f^{[q,s]} (x) + \mathcal{O}(x^{q+s+1})\;.
\end{equation}
Without loss of generality, we take $b_0 = 1$, which effectively reduces the number of unknown coefficients to $q+s+1$. 
Then, all numerator and denominator coefficients, $a_i$ and $b_j$, respectively, are iteratively calculated imposing the correspondence between the Padé approximant of $f(x)$ and its power-series representation.
Hence, Padé approximants capture the behavior of the original function up to $\mathcal{O}(x^{q+s+1})$. 
In particular, we have the freedom to adjust the coefficients $q$ and $s$ to obtain the best fit to the resummed series.

\begin{figure}[h]
\centering
\includegraphics[width=\linewidth]{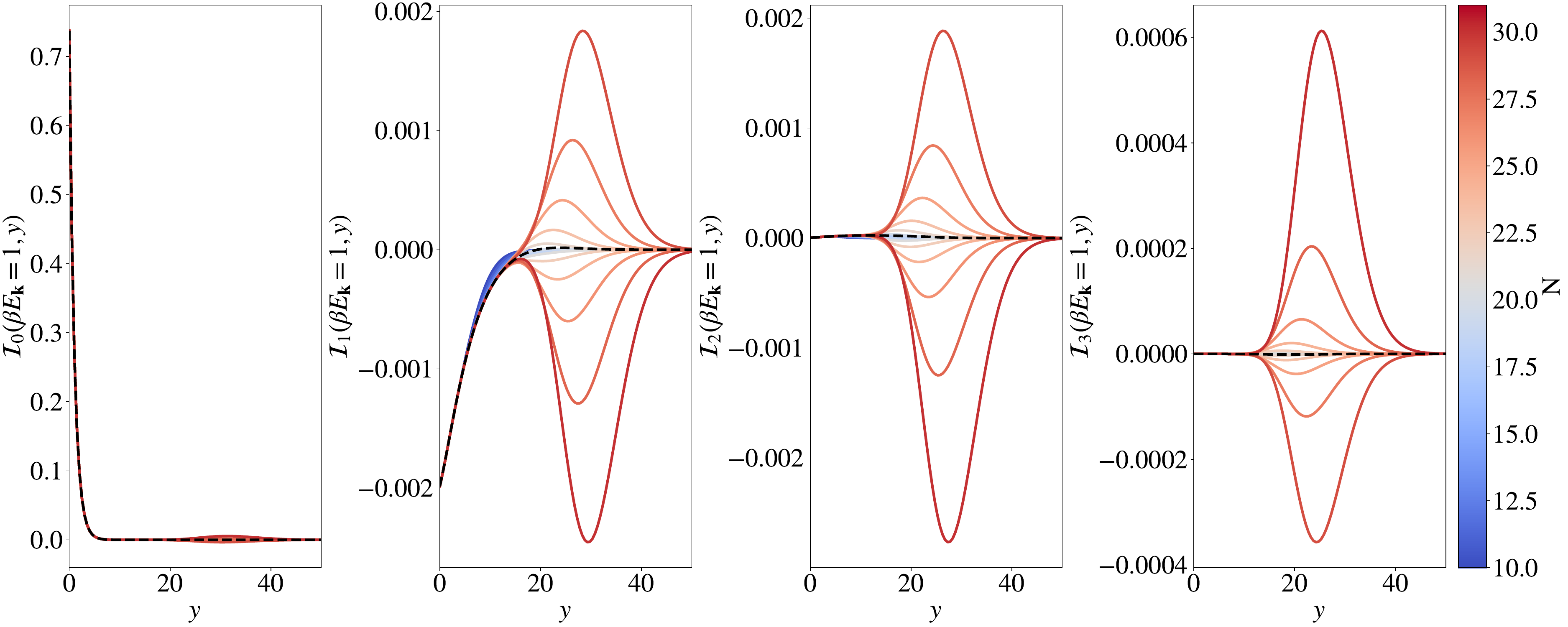}
\caption{Integrand of the Borel-transformed series of different multipoles for a wide range of truncations $N$, considering $\beta E_{\mathbf{k}} = 1$, $\hat{\tau}_{0} = 0.5$ and $\hat{\tau}=10$. The black dashed lines correspond to the Borel-transformed series computed using Padé approximants for the maximal truncation considered here.}
\label{fig:integrand-borel-pade}
\end{figure}

We then compute $\mathcal{I}_{\ell}( \beta E_{\mathbf{k}}, y )$ using Padé approximants with a quadratic polynomial in the denominator. 
In Fig.~\ref{fig:integrand-borel-pade}, we portray $\mathcal{I}_{\ell}( \beta E_{\mathbf{k}}=1, y)$ as a function of $y$ for a wide range of different truncations of the sum in Eq.~\eqref{eq:integrand-borel}, computed using Padé approximants. 
In this case, it can be straightforwardly seen that the Padé approximants (black dashed lines) seem to capture the optimal truncation of the series.

\section{Analytical approach to the origin of the divergence}
\label{sec:analytical}

In this section, we provide an analytical calculation to clarify the origin of the divergence. In particular, we show that the moments $\chi_{n,\ell}$ grow factorially at intermediate times, and that the series for the distribution function becomes (and stays) asymptotic after a certain time.

\subsection{Evolution of the irreducible moments}
\begin{figure}
        \centering
        \includegraphics[scale=0.95]{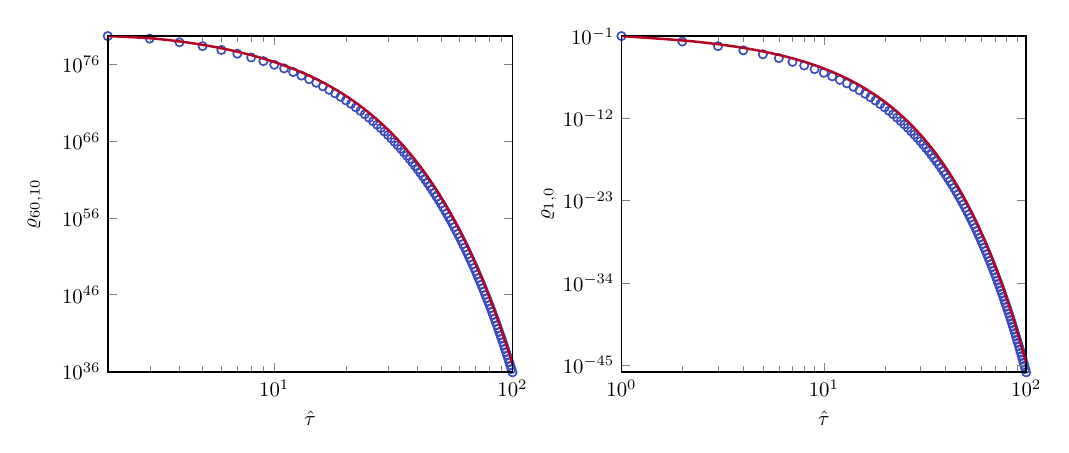}
        \caption{Evolution of different moments (blue circles) in comparison to the analytical approximation (red lines). 
        Left: $(m=60,\ell=10)$, right: $(m=1,\ell=0)$. 
        The chemical potential and the initial temperature are taken as $\alpha=0$ and $T_0=1$. 
        The initial time is taken to be $\hat{\tau}_0=1$.}
        \label{fig:moments_approx}
    \end{figure}

We start from the original equations for the irreducible moments, which take the form \cite{Denicol:2021, deBrito:2024vhm}
\begin{eqnarray}
\partial_\tau \varrho_{m, \ell} & = & - \frac{1}{\tau_R} \left( \varrho_{m, \ell} - \varrho_{m, \ell}^{\mathrm{eq}} \right) - \mathcal{P}(m,\ell) \frac{\varrho_{m, \ell - 1}}{\tau} - \mathcal{Q}(m,\ell)  \frac{\varrho_{m, \ell}}{\tau} - \mathcal{R}(m,\ell)  \frac{\varrho_{m, \ell + 1}}{\tau}\;, \label{eq:eoms-moms}
\end{eqnarray}
where we have introduced the following coefficients
\begin{subequations}
\label{eq:coeffs-eqs-moms}
\begin{align}
\mathcal{P}(m, \ell) &= 2 \ell \frac{(m + 2\ell ) (2\ell - 1)}{(4\ell + 1) (4\ell -1 )}\;, \\
\mathcal{Q}(m, \ell) &=  \frac{2 \ell (2\ell + 1) + m (24\ell^2 + 12 \ell - 3)}{3 (4\ell - 1) (4\ell + 3)} + \frac{2}{3}\;, \\
\mathcal{R}(m, \ell) &= (m - 2 \ell -1) \frac{(2\ell + 1) (2\ell + 2)}{(4\ell + 1) (4\ell + 3)}\;.
\end{align}
\end{subequations} 
Note that the quantity $\mathcal{Q}(m,\ell)$ is slightly different from the coefficient $\mathcal{Q}_{m,\ell}$ used in the main text.
Introducing the vectors
\begin{equation}
    \Vec{\varrho}_m \coloneqq \{\varrho_{m,0},\varrho_{m,1},\cdots\}\;,\qquad \Vec{\varrho}^{\,\text{eq}}_m \coloneqq \{\varrho^{\text{eq}}_{m,0},\varrho^{\text{eq}}_{m,1},\cdots\}\;,
\end{equation}
we can rewrite the equations of motion as
\begin{equation}
    \partial_\tau \vec{\varrho}_m+ \mathbb{A}(m) \vec{\varrho}_m = \frac{1}{\tau_R}\vec{\varrho}_m^{\,\text{eq}} \;, \label{eq:rho_vector}
\end{equation}
where we defined the matrix
\begin{equation}
    \left[\mathbb{A}(m)\right]_{\ell k}\coloneqq \left[\frac{1}{\tau_R}+ \frac{\mathcal{Q}(m,\ell)}{\tau}\right]\delta_{\ell k} + \frac{\mathcal{P}(m,\ell)}{\tau}\delta_{\ell-1,k}+ \frac{\mathcal{R}(m,\ell)}{\tau}\delta_{\ell+1,k}\;.
\end{equation}
We now make several assumptions to simplify the system. First, the equilibrium term on the right-hand side of Eq.~\eqref{eq:rho_vector} provides an asymptotic value for the irreducible moments, and is not expected to be the cause of the pathological behavior. Thus, in the following, we will set it to zero, causing the irreducible moments to relax to zero at asymptotically long times. Furthermore, we take $\tau_R=\mathrm{const}$. In summary, we consider the system of first-order ODEs
\begin{equation}
    \partial_{\hat{\tau}} \vec{\varrho}_m+\left[\mathds{1}+ \frac{1}{\hat{\tau}}\mathbb{B}(m)\right] \vec{\varrho}_m = 0 \;, \label{eq:rho_vector_simple}
\end{equation}
where
\begin{equation}
    \left[\mathbb{B}(m)\right]_{\ell k}\coloneqq  \mathcal{Q}(m,\ell)\delta_{\ell k} + \mathcal{P}(m,\ell)\delta_{\ell-1,k}+ \mathcal{R}(m,\ell)\delta_{\ell+1,k}\;,
\end{equation}
and $\hat{\tau}\coloneqq \tau/\tau_R$, as in the main text.
The system \eqref{eq:rho_vector_simple} is straightforwardly solved by
\begin{equation}
    \vec{\varrho}_m(\hat{\tau})=e^{-(\hat{\tau}-\hat{\tau}_0)}\exp\left[-\ln \left(\hat{\tau}/\hat{\tau}_0\right)\mathbb{B}(m)\right] \vec{\varrho}_m(\hat{\tau}_0)\;.\label{eq:rho_sol}
\end{equation}
In order to proceed, we have to simplify the matrix $\mathbb{B}(m)$. 
Since we observe that the divergences happen for large $m$ at any $\ell>0$, let us first assume that $m\gg 1$,
    \begin{align}
        \mathcal{P}(m,\ell)&=m \frac{2\ell (2\ell-1)}{(4\ell+1)(4\ell-1)}+ \mathcal{O}(1) \;,\\ \mathcal{Q}(m,\ell)&=m \frac{8\ell^2+4\ell-1}{(4\ell-1)(4\ell+3)}+ \mathcal{O}(1)\;,\\
        \mathcal{R}(m,\ell)&=m \frac{(2\ell+1) (2\ell+2)}{(4\ell+1)(4\ell+3)}+ \mathcal{O}(1)\;.
    \end{align}
    Then, the respective moments evolve as
    \begin{equation}
    \vec{\varrho}_m(\hat{\tau})\simeq e^{-(\hat{\tau}-\hat{\tau}_0)}\exp\left[\ln \left(\hat{\tau}_0^m/\hat{\tau}^m\right)\mathbb{D}\right] \vec{\varrho}_m(\hat{\tau}_0)\;,\label{eq:rho_sol_largem}
    \end{equation}
    where we defined
    \begin{equation}
        \mathbb{D}_{\ell k}\coloneqq  \frac{8\ell^2+4\ell-1}{(4\ell-1)(4\ell+3)} \delta_{\ell k} + \frac{2\ell(2\ell-1)}{(4\ell-1)(4\ell+1)}\delta_{\ell-1,k}+\frac{(2\ell+1)(2\ell+2)}{(4\ell+1)(4\ell+3)}\delta_{\ell+1,k}\;,
    \end{equation}
    which is now independent of $m$. 
    To further simplify this matrix, let us assume that $4\ell\gg 3$, such that we can approximate
     \begin{equation}
        \mathbb{D}_{\ell}\simeq  \frac12 \delta_{\ell k} + \frac14\left(\delta_{\ell-1,k}+\delta_{\ell+1,k}\right)\;.
    \end{equation}
    This assumption obviously fails for $\ell=0,1$, but becomes better quickly. We have now reduced the problem to finding the exponential of a tridiagonal symmetric Toeplitz matrix, which has been investigated in Ref.~\cite{tatari2020exponential}. Such an exponential can be approximated in the following manner,
    \begin{equation}
        \left[\exp \begin{pmatrix}
            b& z & 0 &0 & \cdots \\
            z & b & z  & 0 &\cdots\\
            0 & z & b & z & \cdots\\
            0& 0& z & b &\cdots\\
            \vdots & \vdots & \vdots & \vdots & \ddots
        \end{pmatrix}\right]_{\ell k} \simeq 
        e^b \left[I_{\ell-k}(2z)-I_{\ell+k+2}(2z)\right]\;,
    \end{equation}
    where $I_n(x)$ denotes the modified Bessel function of the first kind.\footnote{Note that, in contrast to the notation of Ref.~\cite{tatari2020exponential}, the indices $\ell$ and $k$ start from $0$.}
    Now, we can approximately express the irreducible moments as
    \begin{equation}
        \vec{\varrho}_m(\hat{\tau})\simeq e^{-(\hat{\tau}-\hat{\tau}_0)}\left(\frac{\hat{\tau}_0}{\hat{\tau}}\right)^{m/2}\mathbb{E}(m) \vec{\varrho}_m(\hat{\tau}_0)\;,
    \end{equation}
    where we defined
    \begin{equation}
        \left[\mathbb{E}(m)\right]_{\ell k}\coloneqq I_{\ell-k}\left(\ln \frac{\hat{\tau}_0^{m/2}}{\hat{\tau}^{m/2}}\right)-I_{\ell+k+2}\left(\ln \frac{\hat{\tau}_0^{m/2}}{\hat{\tau}^{m/2}}\right)\;.
    \end{equation}
    Since we initialize the system in thermal equilibrium,
    \begin{equation}
    \varrho_{m,\ell}(\hat{\tau}_0)=\left[\vec{\varrho}_m\right]_\ell(\hat{\tau}_0)= e^\alpha T_0^{m+2} \frac{(m+1)!}{2\pi^2 }\delta_{\ell 0}\;,
    \end{equation}
    we arrive at the result
    \begin{align}
        \varrho_{m,\ell}(\hat{\tau})&\simeq e^\alpha T_0^{m+2} \frac{(m+1)!}{2\pi^2 } e^{-(\hat{\tau}-\hat{\tau}_0)}\left(\frac{\hat{\tau}_0}{\hat{\tau}}\right)^{m/2}\left[I_{\ell}\left(\ln \frac{\hat{\tau}_0^{m/2}}{\hat{\tau}^{m/2}}\right)-I_{\ell+2}\left(\ln \frac{\hat{\tau}_0^{m/2}}{\hat{\tau}^{m/2}}\right)\right]\nonumber\\
        &= e^\alpha T_0^{m+2} \frac{(m+1)!}{2\pi^2 } e^{-(\hat{\tau}-\hat{\tau}_0)}\left(\frac{\hat{\tau}_0}{\hat{\tau}}\right)^{m/2}\frac{4(\ell+1)(-1)^\ell}{m\ln\left(\hat{\tau}/\hat{\tau}_0\right)}I_{\ell+1}\left(\ln \frac{\hat{\tau}^{m/2}}{\hat{\tau}_0^{m/2}}\right)\;.
        \label{eq:sol_rho_final}
    \end{align}
    When comparing Eq.~\eqref{eq:sol_rho_final} to the exact solution \eqref{eq:rho_sol} of the simplified system, it becomes apparent that, even for moderately large $m$ and $\ell\geq 1$, it works rather well, see Fig.~\ref{fig:moments_approx}. Quantitatively, compared to the exact solution there are relative factors of orders $\mathcal{O}(1)-\mathcal{O}(10)$. Nevertheless, the qualitative behavior is captured.

    \subsection{Factorial divergence of the normalized moments}
\begin{figure}
        \centering
        \includegraphics[scale=0.95]{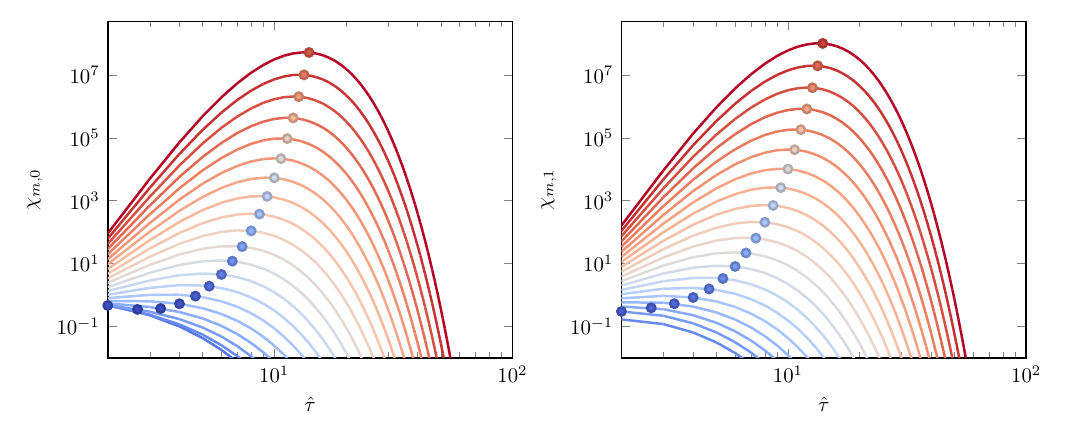}
        \caption{The analytic evolution of the normalized moments $\chi_{m,0}$ and $\chi_{m,1}$ for $m=2,4,\ldots{},40$. It can be clearly seen that they diverge with increasing $m$. The points mark the value of $\chi_{m,\ell}$ at the time $\hat{\tau}=(m+2)/3$. The initial time is taken to be $\hat{\tau}_0=1$.}
        \label{fig:chi_div}
    \end{figure}

    In order to move on to the normalized moments $\chi_{m,\ell}$, we have to first estimate the equilibrium moments. To do this, we consider the equation of motion for the temperature 
    \begin{equation}
    \label{eq:eom-temp}
    \partial_\tau T + \frac{T}{3 \tau} (1 + 2 \chi_{2,1}) = 0\;,
    \end{equation}
    in the limit where it decouples from the dissipative degrees of freedom, i.e., where the term proportional to $\chi_{2,1}$ can be ignored. 
    Then, it is straightforwardly solved as
    \begin{equation}
        T(\hat{\tau})\simeq T_0 \left(\frac{\hat{\tau}_0}{\hat{\tau}}\right)^{1/3}\;,
    \end{equation}
    which gives the equilibrium moments as
    \begin{equation}
        \varrho_{m,0}^{\mathrm{eq}}(\hat{\tau})\simeq e^\alpha \frac{(m+1)!}{2\pi^2} T_0^{m+2} \left(\frac{\hat{\tau}_0}{\hat{\tau}}\right)^{(m+2)/3}\;,\label{eq:rho_eq_approx}
    \end{equation}
    leading us to the normalized moments
    \begin{equation}
        \chi_{m,\ell}(\hat{\tau})=\frac{\varrho_{m,\ell}(\hat{\tau})}{\varrho_{m,0}^{\mathrm{eq}}(\hat{\tau})}\simeq  e^{-(\hat{\tau}-\hat{\tau}_0)}\left(\frac{\hat{\tau}_0}{\hat{\tau}}\right)^{(m-4)/6}\frac{4(\ell+1)(-1)^\ell}{m\ln\left(\hat{\tau}/\hat{\tau}_0\right)}I_{\ell+1}\left(\ln \frac{\hat{\tau}^{m/2}}{\hat{\tau}_0^{m/2}}\right)\;.
        \label{eq:sol_chi_final}
    \end{equation}
    As one can see in Fig.~\ref{fig:chi_div}, we can reproduce the essential features of the growth of $\chi_{m,\ell}$ at intermediate times.
    Furthermore, we can obtain an estimate of the point at which the curve peaks. Using the expansion of the Bessel function for large arguments,
    \begin{equation}
    I_\nu(x)\simeq \frac{e^x}{\sqrt{2\pi x}}\;,\label{eq:Bessel_asymptotic}
    \end{equation}
    which is justified at not too small times since $m$ is large, we can approximate that, up to logarithmic corrections, the rescaled moments  behave as
    \begin{equation}
        \chi_{m,\ell}(\hat{\tau})\sim e^{-(\hat{\tau}-\hat{\tau}_0)} \left(\frac{\hat{\tau}}{\hat{\tau}_0}\right)^{(m+2)/3}\;.\label{eq:chi_approx}
    \end{equation}
    This suggests that the maximum of $\chi_{m,\ell}(\hat{\tau})$ lies at $\hat{\tau}_{\mathrm{max}}/\hat{\tau}_0\simeq (m+2)/3 $, cf. Fig.~\ref{fig:chi_div}. We can then estimate that the  magnitude of the maximum grows as
    \begin{equation}
        \chi_{m,\ell}(\hat{\tau}_{\mathrm{max}})\sim \left(\frac{m/3}{e}\right)^{m/3} \sim (m/3)!\;,
    \end{equation}    
    showing a factorial divergence. 

    \subsection{Summing the distribution function}
    Using the results of the previous section, we are now in a position to analyze what the growth of the moments $\chi_{m,\ell}$ implies for the distribution function. 
    First we compute the coefficients
    \begin{align}
        c_{n,\ell}(\hat{\tau})&=(4\ell+1)  n! \sum_{m=0}^n \frac{(-1)^m \, (m + 2\ell + 1)!}{(n-m)! (m + 4\ell + 1)! m!} \chi_{m + 2\ell, \ell}(\hat{\tau})\nonumber\\
        &=(4\ell+1)4(\ell+1) (-1)^\ell e^{-(t-t_0)} \left(\frac{\hat{\tau}}{\hat{\tau}_0}\right)^{2/3} \frac{1}{\ln\left(\hat{\tau}/\hat{\tau}_0\right)} \nonumber\\
        &\quad\times \sum_{m=0}^n \binom{n}{m} (-1)^m \frac{(m + 2\ell + 1)!}{(m + 4\ell + 1)!} \left(\frac{\hat{\tau}_0}{\hat{\tau}}\right)^{(m+2\ell)/6}\frac{1}{m+2\ell}I_{\ell+1}\left(\frac{m+2\ell}{2}\ln \frac{\hat{\tau}}{\hat{\tau}_0}\right)\;.
    \end{align}
    Since the $\ell$-th multipole moment of the distribution function (truncated at order $N$, which we later send to infinity) is given by
    \begin{equation}
    \mathcal{F}^{(\ell)}_\mathbf{k}(N,\hat{\tau})\coloneqq \frac{2}{4\ell+1}f_{0\mathbf{k}} (\beta E_\mathbf{k})^{2\ell} \sum_{n=0}^N L_n^{(4\ell+1)}(\beta E_\mathbf{k}) c_{n,\ell}(\hat{\tau})\;,\label{eq:f_ell}
    \end{equation}
    we need to evaluate the following expression,
      \begin{align}
    \sum_{n=0}^N L_n^{(4\ell+1)}(\beta E_\mathbf{k}) c_{n,\ell}(\hat{\tau})&=(4\ell+1)4(\ell+1) (-1)^\ell e^{-(\hat{\tau}-\hat{\tau}_0)} \left(\frac{\hat{\tau}}{\hat{\tau}_0}\right)^{(2-\ell)/3} \frac{1}{\ln\left(\hat{\tau}/\hat{\tau}_0\right)}\nonumber\\
    &\; \times\sum_{n=0}^N L_n^{(4\ell+1)}(\beta E_\mathbf{k}) \sum_{m=0}^n \binom{n}{m}\left(-\sqrt[6]{\frac{\hat{\tau}_0}{\hat{\tau}}}\right)^m \frac{(m+2\ell+1)!}{(m+4\ell+1)!} \frac{1}{m+2\ell} I_{\ell+1}\left(\frac{m+2\ell}{2}\ln \frac{\hat{\tau}}{\hat{\tau}_0}\right)\nonumber\\
        &=(4\ell+1)4(\ell+1) (-1)^\ell e^{-(\hat{\tau}-\hat{\tau}_0)} \left(\frac{\hat{\tau}}{\hat{\tau}_0}\right)^{(2-\ell)/3} \frac{1}{\ln\left(\hat{\tau}/\hat{\tau}_0\right)}\nonumber\\
        &\; \times\sum_{m=0}^N \frac{\left(-\sqrt[6]{\hat{\tau}_0/\hat{\tau}}\right)^m}{m!} \frac{(m+2\ell+1)!}{(m+4\ell+1)!} \frac{1}{m+2\ell} I_{\ell+1}\left(\frac{m+2\ell}{2}\ln \frac{\hat{\tau}_0}{\hat{\tau}}\right) \sum_{n=m}^N \frac{n!}{(n-m)!} L_n^{(4\ell+1)}(\beta E_\mathbf{k})\;. 
    \end{align}
    With the help of the integral representation of the associated Laguerre polynomials
    \begin{equation}
        L_n^{(\alpha)}(x)=e^x x^{-\alpha/2} \frac{1}{n!} \int_0^\infty \mathrm{d} s\, e^{-s} s^{n+\alpha/2} J_\alpha\left(2 \sqrt{s x}\right)\;,\label{eq:rep_L}
    \end{equation}
    we are able to compute
    \begin{align}
        \sum_{n=m}^N \frac{n!}{(n-m)!} L_n^{(4\ell+1)}(\beta E_\mathbf{k})
        &=e^{\beta E_\mathbf{k}} (\beta E_\mathbf{k})^{-2\ell-1/2} \int_0^\infty \mathrm{d}s\, e^{-s} s^{2\ell+1/2} J_{4\ell+1}(2\sqrt{s\beta E_\mathbf{k}}) \sum_{n=m}^N \frac{s^n}{(n-m)!}\nonumber\\
        &=e^{\beta E_\mathbf{k}} \frac{(\beta E_\mathbf{k})^{-2\ell-1/2} }{(N-m)!}\int_0^\infty \mathrm{d}s \,s^{2\ell+1/2+m} J_{4\ell+1}(2\sqrt{s\beta E_\mathbf{k}}) \Gamma(N-m+1,s)\;.
    \end{align}
    Writing the incomplete Gamma function as
    \begin{equation}
       \Gamma (n,a) = a^n\int_1^\infty \mathrm{d}u \, e^{-a u} u^{n-1}\;,
    \end{equation}
    and switching the order of integration, we find
    \begin{align}
    \sum_{n=m}^N \frac{n!}{(n-m)!} L_n^{(4\ell+1)}(\beta E_\mathbf{k})
    &=e^{\beta E_\mathbf{k}} \frac{(\beta E_\mathbf{k})^{-2\ell-1/2} }{(N-m)!}
    \int_1^\infty \mathrm{d}u\, u^{N-m}\int_0^\infty \mathrm{d}s \,e^{-su}s^{2\ell+3/2+N} J_{4\ell+1}(2\sqrt{s\beta E_\mathbf{k}}) \nonumber\\
    &=e^{\beta E_\mathbf{k}} \frac{(N+1)!}{(N-m)!}
    \int_0^{1} \mathrm{d}v \,e^{-v\beta E_\mathbf{k}} v^{m+4\ell+1} L_{N+1}^{(4\ell+1)}(v \beta E_\mathbf{k})\nonumber\\
    &=e^{\beta E_\mathbf{k}} \frac{(m+4\ell+1)!}{(4\ell+1)!} \binom{N+4\ell+2}{N-m}\pFq{2}{2}{N+4\ell+3,m+4\ell+2}{4\ell+2,m+4\ell+3}{-\beta E_\mathbf{k}} \;, \label{eq:sum_Laguerre}
    \end{align}
    where we employed Eq.~\eqref{eq:rep_L} and substituted $v\coloneqq 1/u$. 
    The hypergeometric function ${}_2F_2$ arises because of the
    connection between Laguerre polynomials and confluent hypergeometric functions
    \begin{equation}
        L_n^{(\alpha)}(x)=\binom{n+\alpha}{n} \pFq{1}{1}{-n}{\alpha+1}{x}=e^x \binom{n+\alpha}{n} \pFq{1}{1}{n+\alpha+1}{\alpha+1}{-x}\;,
    \end{equation}
    and the integral identity \cite{NIST:DLMF}
    \begin{equation}
        \pFq{2}{2}{a,b}{c,d}{x}=\frac{(d-1)!}{(b-1)!(d-b)!}\int_0^1 \mathrm{d} v \,v^{b-1}(1-v)^{d-b-1} \pFq{1}{1}{a}{c}{xv} \;,\label{eq:int_2F2}
    \end{equation}
    which can be combined (with $a\equiv N+4\ell+3$, $b\equiv m+4\ell+2$, and $c\equiv 4\ell+2$) to yield
    \begin{equation}
        \pFq{2}{2}{a,b}{c,b+1}{-\beta E_\mathbf{k}}= \frac{b(a-c)!(c-1)!}{(a-1)!}
    \int_0^{1} \mathrm{d}v \,e^{-v\beta E_\mathbf{k}} v^{b-1} L_{a-c}^{(c-1)}(v \beta E_\mathbf{k})\;.  \label{eq:connection_F_L}
    \end{equation}
    Inserting the result \eqref{eq:sum_Laguerre} into Eq.~\eqref{eq:f_ell}, we arrive at
    \begin{multline}
    \mathcal{F}^{(\ell)}_\mathbf{k}(N,\hat{\tau})=8e^{\alpha}(\ell+1) (-1)^\ell e^{-(\hat{\tau}-\hat{\tau}_0)} \left(\frac{\hat{\tau}}{\hat{\tau}_0}\right)^{(2-\ell)/3} \frac{(\beta E_\mathbf{k})^{2\ell}}{\ln\left(\hat{\tau}/\hat{\tau}_0\right)}\frac{(N+4\ell+2)!}{(4\ell+1)!N!}\nonumber\\
    \times\sum_{m=0}^N \binom{N}{m} \left(-\sqrt[6]{\frac{\hat{\tau}_0}{\hat{\tau}}}\right)^m\frac{(m+2\ell+1)!}{(m+4\ell+2)!} \frac{I_{\ell+1}\left(\frac{m+2\ell}{2}\ln \frac{\hat{\tau}}{\hat{\tau}_0}\right)}{m+2\ell}    \pFq{2}{2}{N+4\ell+3,m+4\ell+2}{4\ell+2,m+4\ell+3}{-\beta E_\mathbf{k}}\;.
    \end{multline}
    By repeatedly using the relation
    \begin{equation}
        \frac{\partial}{\partial x} \pFq{2}{2}{a,b}{c,d}{x}= \frac{ab}{cd}\pFq{2}{2}{a+1,b+1}{c+1,d+1}{x}\;,
    \end{equation}
    we can rewrite $\mathcal{F}^{(\ell)}_\mathbf{k}$ as
    \begin{align}
    \mathcal{F}^{(\ell)}_\mathbf{k}(N,\hat{\tau})&=8e^{\alpha}(\ell+1) (-1)^\ell e^{-(\hat{\tau}-\hat{\tau}_0)} \left(\frac{\hat{\tau}}{\hat{\tau}_0}\right)^{(2-\ell)/3} \frac{(\beta E_\mathbf{k})^{2\ell}}{\ln\left(\hat{\tau}/\hat{\tau}_0\right)}(N+1)(N+2) \nonumber\\
    &\times \frac{\partial^{4\ell} }{\partial (\beta E_\mathbf{k})^{4\ell}} \sum_{m=0}^N \binom{N}{m}\left(-\sqrt[6]{\frac{\hat{\tau}_0}{\hat{\tau}}}\right)^m\frac{(m+2\ell+1)!}{(m+4\ell+1)!} \frac{I_{\ell+1}\left(\frac{m+2\ell}{2}\ln \frac{\hat{\tau}}{\hat{\tau}_0}\right)}{(m+2\ell)(m+2)} \pFq{2}{2}{N+3,m+2}{2,m+3}{-\beta E_\mathbf{k}}\;.\label{eq:f_ell_full}
    \end{align}
    The crucial ingredient is the sum that appears in the second line of the equation above, which we will analyze more closely in the following.
    Using the integral representation of the Bessel function
    \begin{equation}
        I_\nu (x)=\frac{(x/2)^\nu}{\sqrt{\pi}\Gamma\left(\nu+\frac12\right)} \int_0^\pi \mathrm{d}\vartheta \, \left(\sin\vartheta\right)^{2\nu} e^{x\cos\vartheta}\;,\label{eq:Bessel_int}
    \end{equation}
    as well as 
    \begin{equation}
        \pFq{2}{2}{N+3, m+2}{2,m+3}{-\beta E_\mathbf{k}}= (m+2) \int_0^1 \mathrm{d}v\, v^{m+1} \pFq{1}{1}{N+3}{2}{-v\beta E_\mathbf{k}}\;,
    \end{equation}
we can rewrite the sum in $\mathcal{F}^{(\ell)}_\mathbf{k}$ as
\begin{align}
    &\quad(N+1)(N+2)\sum_{m=0}^N \binom{N}{m}\left(-\sqrt[6]{\frac{\hat{\tau}_0}{\hat{\tau}}}\right)^m\frac{(m+2\ell+1)!}{(m+4\ell+1)!} \frac{I_{\ell+1}\left(\frac{m+2\ell}{2}\ln \frac{\hat{\tau}}{\hat{\tau}_0}\right)}{(m+2\ell)(m+2)} \pFq{2}{2}{N+3,m+2}{2,m+3}{-\beta E_\mathbf{k}}\nonumber\\
    &=(N+1)(N+2)\frac{\left(\ln \hat{\tau}/\hat{\tau}_0\right)^{\ell+1}}{4^{\ell+1}\sqrt{\pi}\Gamma\left(\ell+\frac32\right)} \int_0^\pi \mathrm{d} \vartheta \left(\sin\vartheta\right)^{2\ell+2}\left(\frac{\hat{\tau}}{\hat{\tau}_0}\right)^{\ell\cos\vartheta}\nonumber\\
    &\qquad \times\sum_{m=0}^N \binom{N}{m}\left[-\left(\frac{\hat{\tau}}{\hat{\tau}_0}\right)^{\cos\vartheta/2-1/6}\right]^m\frac{(m+2\ell+1)!}{(m+4\ell+1)!} \frac{(m+2\ell)^\ell}{m+2} \pFq{2}{2}{N+3,m+2}{2,m+3}{-\beta E_\mathbf{k}}\nonumber\\
    &=(N+1)(N+2)\frac{\left(\ln \hat{\tau}/\hat{\tau}_0\right)^{\ell+1}}{4^{\ell+1}\sqrt{\pi}\Gamma\left(\ell+\frac32\right)} \int_0^\pi \mathrm{d} \vartheta \left(\sin\vartheta\right)^{2\ell+2}\left(\frac{\hat{\tau}}{\hat{\tau}_0}\right)^{\ell\cos\vartheta} \int_0^1 \mathrm{d} v\, v \pFq{1}{1}{N+3}{2}{-v\beta E_\mathbf{k}}\nonumber\\
    &\qquad \times\sum_{m=0}^N \binom{N}{m}\left[-v\left(\frac{\hat{\tau}}{\hat{\tau}_0}\right)^{\cos\vartheta/2-1/6}\right]^m\frac{(m+2\ell+1)!}{(m+4\ell+1)!} (m+2\ell)^\ell\;.\label{eq:f_sum_l}
    \end{align}
    In the following, we will first specialize to the case $\ell=0$, and subsequently treat the more involved case $\ell>0$, which works in a similar way, but is technically more complicated.
    
    \subsubsection{Case \texorpdfstring{$\ell=0$}{l=0}}
    From Eq.~\eqref{eq:f_sum_l}, we obtain in the case where $\ell=0$,
    \begin{align}
    &\quad(N+1)(N+2)\sum_{m=0}^N \binom{N}{m}\left(-\sqrt[6]{\frac{\hat{\tau}_0}{\hat{\tau}}}\right)^m\frac{I_{1}\left(\frac{m}{2}\ln \frac{\hat{\tau}}{\hat{\tau}_0}\right)}{m(m+2)} \pFq{2}{2}{N+3,m+2}{2,m+3}{-\beta E_\mathbf{k}}\nonumber\\
    &=(N+1)(N+2)\frac{\ln \hat{\tau}/\hat{\tau}_0}{2\pi} \int_0^\pi \mathrm{d} \vartheta \sin^2\vartheta \int_0^1 \mathrm{d} v\, v \pFq{1}{1}{N+3}{2}{-v\beta E_\mathbf{k}}\left[1-v y(\vartheta)\right]^N\nonumber\\
    &=-(N+2)\frac{\ln \hat{\tau}/\hat{\tau}_0}{2\pi} \int_0^\pi \mathrm{d} \vartheta \sin^2\vartheta \frac{\partial}{\partial y(\vartheta)}\int_0^1 \mathrm{d} v\, \pFq{1}{1}{N+3}{2}{-v\beta E_\mathbf{k}}\left[1-v y(\vartheta)\right]^{N+1}\;, \label{eq:sumint}
    \end{align}
    where we defined $y(\vartheta)\coloneqq \left(\hat{\tau}/\hat{\tau}_0\right)^{\cos\vartheta/2-1/6}$.
    By looking at the integrand in Eq.~\eqref{eq:sumint}, we can identify the crucial ingredient for the observed divergence of the distribution function: as long as the integration interval is such that $y\in(0,2)$ always holds, the integrand converges for $N\to\infty$. Conversely, if $y\leq 0$ or $y\geq 2$ at some point inside the integral, the limit $N\to\infty$ does not exist, implying that the series for the distribution function becomes asymptotic. 

    At the upper limit of the $\vartheta$-integration, we have $y(\pi)=(\hat{\tau}/\hat{\tau}_0)^{-2/3}$, which converges to zero from above for large times and is thus well behaved. 
    Conversely, at the lower limit of the $\vartheta$-integration, we have 
    $y(0)=(\hat{\tau}/\hat{\tau}_0)^{1/3}$, which leads to divergences as soon as $\hat{\tau}/\hat{\tau}_0> 8$. 
    The edge case $\hat{\tau}/\hat{\tau}_0=8$ is still well behaved, since then the divergence only occurs at $\vartheta=0$, and that point carries zero weight due to the factors of $\sin\vartheta$.
    Thus, we can conclude that the series for the distribution function is asymptotic for times $\hat{\tau}/\hat{\tau}_0>8$.

    In the following, we will assume that $\hat{\tau}/\hat{\tau}_0\leq 8$, such that the limit $N\to \infty$ exists.
    In order to perform the $v$-integral, we note that for $N\gg 1$
    \begin{align}
        \pFq{1}{1}{N+3}{2}{-x}&= \sum_{k=0}^\infty \frac{(-x)^k}{k!} \frac{(N+3)_k}{(2)_k}\nonumber\\
        &\approx \sum_{k=0}^\infty \frac{\left[-(N+2)x\right]^k}{k!} \frac{1}{(2)_k} = \pFq{0}{1}{}{2}{-(N+2)x}= \frac{J_1\left(2\sqrt{(N+2)x}\right)}{\sqrt{(N+2)x}}\;.
    \end{align}
    Using this relation and substituting $u\coloneqq 2 \sqrt{(N+2)v \beta E_\mathbf{k}}$, Eq.~\eqref{eq:sumint} becomes
    \begin{align}
        &\quad-(N+2)\frac{\ln \hat{\tau}/\hat{\tau}_0}{2\pi} \int_0^\pi \mathrm{d} \vartheta \sin^2\vartheta \frac{\partial}{\partial y(\vartheta)}\int_0^1 \mathrm{d} v\, \pFq{1}{1}{N+3}{2}{-v\beta E_\mathbf{k}}\left[1-v y(\vartheta)\right]^{N+1}\nonumber\\
        &=-\frac{\ln \hat{\tau}/\hat{\tau}_0}{2\pi \beta E_\mathbf{k}} \int_0^\pi \mathrm{d} \vartheta \sin^2\vartheta \frac{\partial}{\partial y(\vartheta)}\int_0^{2\sqrt{(N+2)\beta E_\mathbf{k}}} \mathrm{d} u\, J_1(u)\left[1-\frac{y(\vartheta)}{4(N+2)\beta E_\mathbf{k}}u^2\right]^{N+1}\nonumber\\
        &\approx-\frac{\ln \hat{\tau}/\hat{\tau}_0}{2\pi \beta E_\mathbf{k}} \int_0^\pi \mathrm{d} \vartheta \sin^2\vartheta \frac{\partial}{\partial y(\vartheta)}\int_0^{\infty} \mathrm{d} u\, J_1(u)\exp\left[-\frac{y(\vartheta)}{4\beta E_\mathbf{k}}u^2\right]\;,
    \end{align}
    where we used that $N\gg 1$ in the second step and expanded $\ln(1-x)\approx -x$ for small $x$. Note that, at this point, all dependence on the truncation $N$ is gone, which is due to the fact that we put the upper limit of integration to infinity [and approximated $(N+1)/(N+2)\approx 1$]. This procedure is permissible as long as $\hat{\tau}/\hat{\tau}_0\leq 8$; for larger values of $\hat{\tau}/\hat{\tau}_0$ it amounts to neglecting an infinite contribution. 
    Employing the integral (cf. Eq.~10.22.54 of Ref.~\cite{NIST:DLMF})
    \begin{equation}
        \int_0^\infty \d u\, J_1(u) e^{-a u^2}= 1-e^{-1/(4a)}\;,
    \end{equation}
    we find 
    \begin{equation}
        -\frac{\ln \hat{\tau}/\hat{\tau}_0}{2\pi \beta E_\mathbf{k}} \int_0^\pi \mathrm{d} \vartheta \sin^2\vartheta \frac{\partial}{\partial y(\vartheta)}\int_0^{\infty} \mathrm{d} u\, J_1(u)\exp\left[-\frac{y(\vartheta)}{4\beta E_\mathbf{k}}u^2\right]=\frac{\ln \hat{\tau}/\hat{\tau}_0}{2\pi } \int_0^\pi \mathrm{d} \vartheta \sin^2\vartheta \frac{\exp\left[-\frac{\beta E_\mathbf{k}}{y(\vartheta)}\right]}{y(\vartheta)^2}\;.
    \end{equation}
    Using the definition of $y(\vartheta)$, the Taylor series of the exponential function, and the integral representation \eqref{eq:Bessel_int} finally gives
    \begin{align}
        &\quad \frac{\ln \hat{\tau}/\hat{\tau}_0}{2\pi } \int_0^\pi \mathrm{d} \vartheta \sin^2\vartheta \frac{\exp\left[-\frac{\beta E_\mathbf{k}}{y(\vartheta)}\right]}{y(\vartheta)^2}\nonumber\\
        &=\left(\frac{\hat{\tau}}{\hat{\tau}_0}\right)^{1/3}\frac{\ln \hat{\tau}/\hat{\tau}_0}{2\pi } \sum_{k=0}^\infty \frac{1}{k!} \left[-\beta E_\mathbf{k} \left(\frac{\hat{\tau}}{\hat{\tau}_0}\right)^{1/6}\right]^k\int_0^\pi \mathrm{d} \vartheta \sin^2\vartheta \exp\left[-\cos\vartheta \ln \frac{\hat{\tau}}{\hat{\tau}_0}\left(1+\frac{k}{2}\right)\right]\nonumber\\
        &=\left(\frac{\hat{\tau}}{\hat{\tau}_0}\right)^{1/3} \sum_{k=0}^\infty \frac{1}{k! (k+2)} \left[-\beta E_\mathbf{k} \left(\frac{\hat{\tau}}{\hat{\tau}_0}\right)^{1/6}\right]^k I_1\left(\frac{k+2}{2}\ln \frac{\hat{\tau}}{\hat{\tau}_0}\right)\;.
    \end{align}
    Thus, making use of Eq.~\eqref{eq:f_ell_full}, we find the zeroth multipole moment of the distribution function to be (for $\hat{\tau}/\hat{\tau}_0\leq 8$)
    \begin{equation}
        \mathcal{F}^{(0)}_\mathbf{k}(\hat{\tau})\equiv \lim_{N\to\infty} \mathcal{F}^{(0)}_\mathbf{k}(N,\hat{\tau})=8 e^\alpha \frac{\hat{\tau}}{\hat{\tau}_0} \frac{e^{-(\hat{\tau}-\hat{\tau}_0)}}{\ln \hat{\tau}/\hat{\tau}_0} \sum_{k=0}^\infty \frac{1}{k! (k+2)} \left[-\beta E_\mathbf{k} \left(\frac{\hat{\tau}}{\hat{\tau}_0}\right)^{1/6}\right]^k I_1\left(\frac{k+2}{2}\ln \frac{\hat{\tau}}{\hat{\tau}_0}\right)\;,\label{eq:f0_final}
    \end{equation}
    whereas the limit does not exist in the standard sense for $\hat{\tau}/\hat{\tau}_0>8$.
    
\begin{figure}
    \centering
    \includegraphics{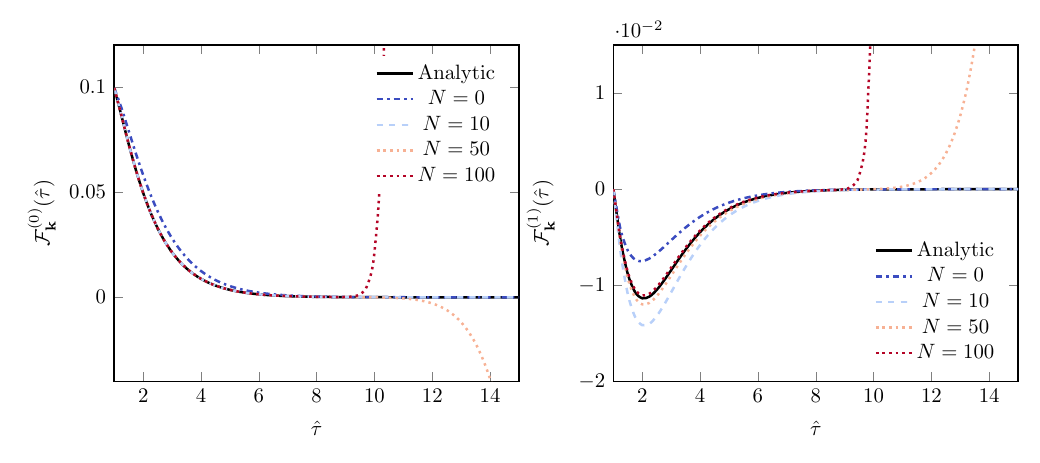}
    \caption{The zeroth (left panel) and first (right panel) multipole moments of the distribution function at energy $\beta E_\mathbf{k}=3$ as functions of time for different values of $N$. 
    The lines for finite $N$ correspond to the evaluation of Eq.~\eqref{eq:f_ell_full}, while the black line denotes the analytical solution given by Eq.~\eqref{eq:fl_final}.  
    In both cases, the initial time is taken to be $\hat{\tau}_0=1$.}
    \label{fig:f_time}
\end{figure}

    In closing, let us note two special cases: First, for $\hat{\tau}/\hat{\tau}_0=1$, we may use the expansion 
    \begin{equation}
        I_\nu(x)\approx \frac{1}{\nu!}\left(\frac{x}{2}\right)^\nu\label{eq:Bessel_approx}
    \end{equation}
    to obtain
    \begin{equation}
        \mathcal{F}^{(0)}_\mathbf{k}(\hat{\tau}_0)= 2e^\alpha \sum_{k=0}^\infty \frac{1}{k!} \left(-\beta E_\mathbf{k} \right)^k = 2e^{\alpha-\beta E_\mathbf{k}}\;,
    \end{equation}
    as expected, since we consider the system to be initially in an equilibrium state.
    Second, for $\beta E_\mathbf{k}=0$, only the $k=0$-term contributes and we have
    \begin{equation}
        \mathcal{F}^{(0)}_\mathbf{0}(\hat{\tau})=4 e^\alpha \frac{\hat{\tau}}{t_0} \frac{e^{-(t-\hat{\tau}_0)}}{\ln \hat{\tau}/\hat{\tau}_0}  I_1\left(\ln \frac{\hat{\tau}}{\hat{\tau}_0}\right)\;.
    \end{equation}
We display the zeroth and first multipoles of the single-particle distribution function for $\beta E_{\mathbf{k}} = 3$ as functions of the rescaled proper time in Fig.~\ref{fig:f_time}. We observe that the convergence of multipoles of higher order requires the inclusion of more terms in the expansion. Moreover, it can be clearly seen that the expansion indeed becomes asymptotic for times $\hat{\tau}/\hat{\tau}_0 > 8$.

    \subsubsection{Case \texorpdfstring{$\ell> 0$}{l>0}}
    The argumentation for $\ell>0$ works in the same way as shown before. 
    The main complication lies in evaluating the sum in Eq.~\eqref{eq:f_sum_l}, which can be done by noting that
\begin{align}
    \sum_{m=0}^{N} \binom{N}{m} (-y)^m \frac{(m+2\ell+1)!}{(m+4\ell+1)!}(m+2\ell)^\ell
    &= \lim_{a\to 1} \frac{\partial^\ell}{\partial\ln a^{\ell}}\sum_{m=0}^{N} \binom{N}{m}   \frac{(-y)^m a^{m+2\ell}}{(m+2\ell+2)\cdots(m+4\ell)(m+4\ell+1)}\nonumber\\
    &=\lim_{a\to 1} \frac{\partial^\ell}{\partial\ln a^{\ell}}a^{-2\ell-1} \int \left(\mathrm{d} a\right)^{2\ell} \sum_{m=0}^{N} \binom{N}{m} (-y)^m  a^{m+2\ell+1} \nonumber\\
    &=\lim_{a\to 1} \frac{\partial^\ell}{\partial\ln a^{\ell}}a^{-2\ell-1} \int \left(\mathrm{d} a\right)^{2\ell}  \, a^{2\ell+1} (1-ay)^{N} \;,
\end{align}
where the notation $\int (\mathrm{d}a)^{2\ell} $ means that one has to integrate $2\ell$ times with respect to $a$, and the limit $a\to 1$ has to be taken at the end. 
From this expression, it is already clear that the fact that the series for the distribution function becomes asymptotic for $\hat{\tau}/\hat{\tau}_0>8$ is not changed.
In order to obtain the finite contribution in the case $\hat{\tau}/\hat{\tau}_0\leq 8$, we can repeat the same steps as outlined above [with $y(\vartheta)$ replaced by $a y(\vartheta)$] and find from Eq.~\eqref{eq:f_ell_full}
\begin{equation}
    \mathcal{F}^{(\ell)}_\mathbf{k}(\hat{\tau})=8e^{\alpha}(\ell+1) (-1)^\ell e^{-(\hat{\tau}-\hat{\tau}_0)} \left(\frac{\hat{\tau}}{\hat{\tau}_0}\right)^{(2-\ell)/3} \frac{(\beta E_\mathbf{k})^{2\ell}}{\ln\left(\hat{\tau}/\hat{\tau}_0\right)}g_\ell\left(\beta E_\mathbf{k},\hat{\tau}\right) \;,\label{eq:fct_f}
\end{equation}
    where we defined
    \begin{align}
        g_\ell\left(\beta E_\mathbf{k},\hat{\tau}\right)&\coloneqq  \frac{\partial^{4\ell} }{\partial (\beta E_\mathbf{k})^{4\ell}} \lim_{a\to 1}\frac{\partial^\ell}{\partial\ln a^{\ell}}a^{-2\ell-1} \int \left(\mathrm{d} a\right)^{2\ell}  \, \frac{a^{2\ell-1} \left(\ln \hat{\tau}/\hat{\tau}_0\right)^{\ell+1}}{4^{\ell+1} \sqrt{\pi} \Gamma\left(\ell+\frac32 \right)} \int_0^\pi \mathrm{d} \vartheta \left(\sin\vartheta\right)^{2\ell+2} \left(\frac{\hat{\tau}}{\hat{\tau}_0}\right)^{\ell\cos\vartheta} \frac{\exp\left[-\frac{\beta E_\mathbf{k}}{a y(\vartheta)}\right]}{y(\vartheta)^2}\nonumber\\
        &= \left(\frac{\hat{\tau}}{\hat{\tau}_0}\right)^{(1+2\ell)/3}\frac{ \left(\ln \hat{\tau}/\hat{\tau}_0\right)^{\ell+1}}{4^{\ell+1} \sqrt{\pi} \Gamma\left(\ell+\frac32 \right)} \int_0^\pi \mathrm{d} \vartheta \left(\sin\vartheta\right)^{2\ell+2} \left(\frac{\hat{\tau}}{\hat{\tau}_0}\right)^{-(\ell+1)\cos\vartheta} 
        h_\ell\left(\beta E_\mathbf{k},\vartheta\right)
        \;.\label{eq:fct_g}
    \end{align}
    The function $h_\ell$ is given by
    \begin{align}
        h_\ell\left(\beta E_\mathbf{k},\vartheta\right)&\coloneqq \lim_{a\to 1}\frac{\partial^\ell}{\partial\ln a^{\ell}}a^{-2\ell-1} \int \left(\mathrm{d} a\right)^{2\ell}  \,a^{-2\ell-1} \exp\left[-\frac{\beta E_\mathbf{k}}{a y(\vartheta)}\right]\nonumber\\
        &=\lim_{a\to 1}\frac{\partial^\ell}{\partial\ln a^{\ell}}\frac{\left[y(\vartheta)\right]^{2\ell}}{a^2\left(\beta E_\mathbf{k}\right)^{2\ell}}\exp\left[-\frac{\beta E_\mathbf{k}}{a y(\vartheta)}\right]\nonumber\\
        &=(-1)^\ell\left[\frac{y(\vartheta)}{\beta E_\mathbf{k}}\right]^{2\ell}\sum_{k=0}^\infty \frac{(k+2)^\ell}{k!} \left[-\frac{\beta E_\mathbf{k}}{ y(\vartheta)}\right]^k \;,\label{eq:fct_h}
    \end{align}
    where we used
    \begin{equation}
        \int \left(\mathrm{d}a\right)^{2\ell} a^{-2\ell-1} e^{-\frac{r}{a}}=\frac{a^{2\ell-1}}{r^{2\ell}} e^{-\frac{r}{a}}\;,
    \end{equation}
    which can be proven by induction.
    Finally, inserting Eq.~\eqref{eq:fct_h} into Eq.~\eqref{eq:fct_g} and subsequently into Eq.~\eqref{eq:fct_f}, we obtain
    \begin{align}
    \mathcal{F}^{(\ell)}_\mathbf{k}(\hat{\tau})&=8e^{\alpha}(\ell+1)  \frac{\hat{\tau}}{\hat{\tau}_0} \frac{e^{-(\hat{\tau}-\hat{\tau}_0)}}{\ln\left(\hat{\tau}/\hat{\tau}_0\right)}\sum_{k=0}^\infty \frac{1}{k!(k+2)} \left[-\beta E_\mathbf{k}\left(\frac{\hat{\tau}}{\hat{\tau}_0}\right)^{1/6}\right]^k I_{\ell+1}\left(\frac{k+2}{2}\ln \frac{\hat{\tau}}{\hat{\tau}_0}\right)\;,\label{eq:fl_final}
\end{align}
    where we made use of the integral representation \eqref{eq:Bessel_int}. 
    When comparing Eq.~\eqref{eq:fl_final} to Eq.~\eqref{eq:f0_final}, we observe that the $\ell$-th multipole moment of the distribution function is very similar to the zeroth one, the main difference being the order of the modified Bessel function.

\begin{figure}
    \centering
    \includegraphics{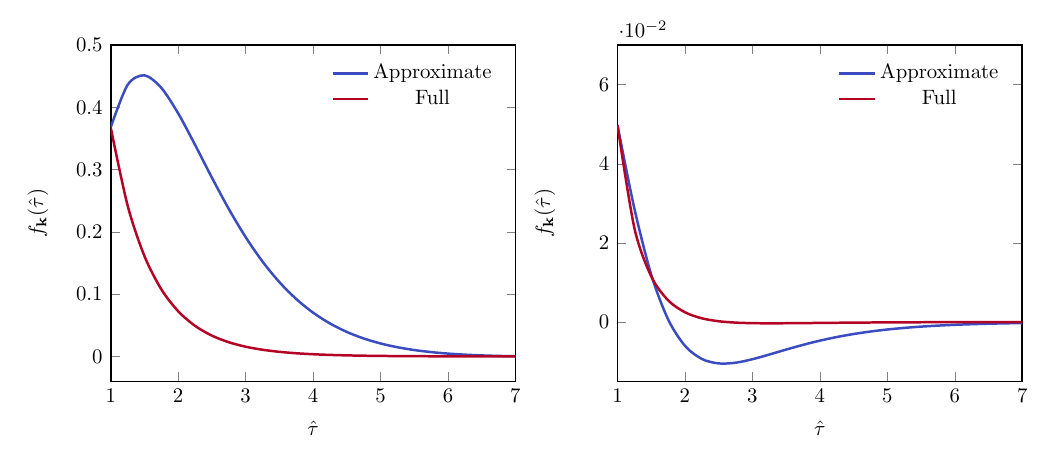}
    \caption{The analytic distribution function computed by summing Eq.~\eqref{eq:fl_final} for $\ell$ up to 4 and the full solution computed with Eq.~\eqref{eq:rho_sol} for $\beta E_\mathbf{k}=1$ (left panel) and $\beta E_\mathbf{k}=3$ (right panel) as functions of time. The initial time is taken to be $\hat{\tau}_0=1$, and we set $\Theta=0$.}
    \label{fig:comparison}
\end{figure}

    Lastly, we have to make a remark on the quantitative accuracy of the distribution function derived here. 
    Although the analytical form obtained in Eq.~\eqref{eq:fl_final} is rather elegant, the quantitative agreement with the full distribution function constructed from the solution \eqref{eq:rho_sol} is not good, as can be seen in Fig.~\ref{fig:comparison}. 
    The reason for this is the approximation made in the solution for the moments \eqref{eq:sol_rho_final}, which captures the important qualitative aspects (such as the expansion becoming asymptotic), but is off by factors of up to order $\mathcal{O}(10)$, which manifests itself in the distribution function. 
    Nevertheless, the calculations performed here shed light on the origin of the observed pathological behavior of the moment expansion.

\end{document}